\documentclass[aps,prb,groupedaddress,twocolumn,floatfix,longbibliography,10pt]{revtex4-2}
\usepackage{amsmath}
\usepackage{amssymb}
\usepackage{graphicx}
\usepackage{bbold}
\usepackage{xcolor}
\usepackage[inkscapelatex=false]{svg}
\usepackage{subcaption}
\usepackage[format=plain,justification=raggedright,singlelinecheck=false]{caption}
\usepackage{times}
\usepackage{hyperref} 
\hypersetup{colorlinks, allcolors=blue}

\allowdisplaybreaks
\begin{document}

\title{Inherent altermagnetism in minimal tight-binding models of regular hyperbolic lattices}
\author{Eric Petermann}
\email{eric.petermann@uni-wuerzburg.de}
\author{Kristian M{\ae}land}
\author{Haye Hinrichsen}
\author{Bj{\"o}rn Trauzettel}
\affiliation{Institute for Theoretical Physics and Astrophysics, University of W{\"u}rzburg, D-97074 W{\"u}rzburg, Germany}
\affiliation{Würzburg-Dresden Cluster of Excellence ctd.qmat, D-97074 W{\"u}rzburg, Germany}

\date{\today}

\begin{abstract}
Altermagnets are a novel class of magnetic systems characterized by their momentum-dependent spin splitting without net magnetization. In this work, we extend established Euclidean tight-binding models of altermagnets to regular hyperbolic lattices in two spatial dimensions defined on a discretized Poincaré disk. Using hyperbolic crystallography and Abelian hyperbolic band theory, we show that the inclusion of next-nearest neighbor hopping is sufficient to induce spin splitting in minimal tight-binding models of bipartite hyperbolic lattices. 
While certain families and special cases of hyperbolic lattices remain antiferromagnetic, we identify an entire family and a special case that generically permit spin splitting in this framework. Hence, altermagnetism is inherent to certain hyperbolic lattices. Since Abelian hyperbolic band theory yields a momentum space that is at least four-dimensional, we classify the leading spin-splitting harmonics using four-dimensional atomic orbitals. As an outlook, we apply non-Abelian hyperbolic band theory to the $\{10,5\}$ lattice. Although its Abelian spectrum is spin degenerate, two selected higher-genus supercells exhibit spin splitting. This finding suggests, that spin degeneracy does not necessarily persist within all translation representation sectors.
\end{abstract}

\maketitle

\def\vec#1{\mathbf{#1}}

\section{Introduction}

Recently, altermagnetism has been proposed as a symmetry class of collinear compensated magnets whose electronic band structures show large spin splitting in the non-relativistic limit \cite{Smejkal2022Sep,Hayami2019Nov,Smejkal2020Jun, Yuan2020Jul, McClarty2024Apr,Bhowal2024Feb}. 
This makes altermagnets promising materials for spintronic applications, as they combine a lattice with negligible stray fields with spin-polarized electronic responses stemming from the momentum-dependent spin splitting \cite{Smejkal2022Dec, Smejkal2022Feb}.
In conventional antiferromagnets, band degeneracy is usually enforced by inversion $\mathcal{P}$ or translations $\tau$ combined with time reversal $\mathcal{T}$. In contrast to this, altermagnets allow for spin-split bands because the magnetic state breaks the degeneracy-enforcing symmetries while preserving combined spin-space and real space operations that enforce degeneracies only along nodal lines. The resulting spin polarization patterns in momentum space are associated with even-parity $d$-, $g$-, or $i$-wave symmetry. This particular magnetic order has motivated various searches for candidate materials, such as $\mathrm{RuO}_2$, MnTe, CrSb, $\mathrm{Mn}_5\mathrm{Si}_3$, and $\mathrm{La}_2\mathrm{O}_3\mathrm{Mn}_2\mathrm{Se}_2$ \cite{Fedchenko2024Jan,Liu2024Oct,Mazin2023Mar,Lee2024Jan,Reimers2024Mar,Ding2024Nov,Reichlova2024Jun,Rial2024Dec,LAOMNSESource,LAOMNSESource2}, as well as measurement methods based on spin- and angle-resolved photoemission, electrical switching and x-ray spectroscopy \cite{Krempasky2024Feb,Han2024Jan,Biniskos2025Oct}. In minimal models of altermagnetism, $\mathcal{PT}$ is typically broken by anisotropic hopping structures \cite{Das2024Jun,Roig2024Oct}, e.g., arising from orbital ordering \cite{Leeb2024Jun}, or by lattice geometries that naturally support the required sublattice asymmetries, as in, e.g., the Lieb lattice \cite{Lieb1989Mar, Brekke2023Dec, MaelandLieb24, Leraand2025Sep,  franzlieblattice,Durrnagel2025JulLieb,Xu2025Lieb,Chang2025Lieb,Petermann2025Dec}.

While altermagnetism has so far been formulated mainly for crystals in flat Euclidean space, regular hyperbolic lattices have emerged as a complementary setting for tight-binding quantum matter on negatively curved manifolds. These systems are not primarily motivated as models of naturally occurring crystals, but rather as potential platforms in which one can test how concepts of lattice quantum mechanics, band theory, topology, and correlation physics translate to non-Euclidean settings \cite{Maciejko2021Sep,Maciejko2022Mar,Boettcher2022Mar, Chen2023Aug}. A regular $2$D hyperbolic lattice is denoted by the Schläfli symbol $\{p,q\}$, which describes a lattice where $q$ $p$-gons meet at each vertex. A hyperbolic lattice must satisfy $(p-2)(q-2)>4$. Hyperbolic lattices have now been realized in circuit quantum electrodynamics, topoelectrical circuits, and photonic resonator arrays \cite{Kollar2019Jul,Zhang2022May,Huang2024Feb}. 

At the single-particle level, hyperbolic lattices support Bloch-like band structures and have already been shown to host hyperbolic topological band insulators, higher-order topological phases, higher-dimensional band topology, and non-Abelian semimetals \cite{Urwyler2022Dec,Liu2023Mar,Zhang2023Feb,Tummuru2024May}.
This makes hyperbolic lattices an interesting setting for investigating symmetry-based magnetic phenomena. For altermagnetism, the central question is whether a compensated collinear magnetic order can generate momentum-dependent spin splitting once Euclidean space-group operations are replaced by hyperbolic point group symmetries and corresponding translations. Moreover, the higher-dimensional momentum space of hyperbolic band theory suggests that the resulting nodal structures and degeneracy manifolds may be richer than in flat space. 

In this work we classify a hyperbolic altermagnet as any lattice with a compensated magnetic order in real space and a spin-split band structure in higher-dimensional momentum space. Specifically, we extend a minimal Euclidean tight-binding model of altermagnetism to bipartite two-dimensional hyperbolic lattices defined on a discretized Poincar\'e disk using Abelian hyperbolic band theory. We show that next-nearest-neighbor hopping alone is sufficient to induce spin splitting in the examined members of the $\{4g,4\}$ family of lattices as well as in the exceptional $\{4,8\}$ lattice, while the $\{2(2g+1),\,2g+1\}$ and $\{2(2g+1),\,3\}$ families and the exceptional $\{8,3\}$ lattice remain spin degenerate. We further analyze the symmetry operations responsible for the remaining degeneracies in the altermagnetic candidates, identify their nodal structures, and compare the resulting momentum-space patterns with those known from Euclidean altermagnets. 

As previously mentioned, our primary focus lies on the Abelian sector of hyperbolic band theory. In this regime, the generators of translations of the band theory commute. This sector does not capture the full set of eigenstates. The remainder of the spectrum is described by non-Abelian hyperbolic band theory \cite{Lenggenhager2023Dec}, which uses larger unit cells in order to access states associated with non-commuting translation generators. 

In Sec.~\ref{sec:theory} to \ref{sec:hyperalt} we focus on the Abelian sector, as it allows us to illustrate our key findings in a pedagogical way. Moreover, the Abelian eigenstates remain valid within the full, non-Abelian description of the problem. However, in Sec.~\ref{sec:nonabelian} we demonstrate, that a non-Abelian extension of our model can give rise to new types of spin splittings.

A similar study has recently examined altermagnetism on hyperbolic lattices \cite{Wang2026Feb}. That work also considers altermagnetism on negatively curved space, but it focuses on truncated, non-regular hyperbolic tilings. Here, we instead give a full symmetry-based analysis of regular hyperbolic tilings. Importantly, we show that regular hyperbolic lattices can host spin split electron bands without hopping anisotropy, indicating that altermagnetism is inherent to numerous hyperbolic lattices.

The article is organized as follows. In Sec.~\ref{sec:theory} we describe how the hyperbolic lattices and the lattice Hamiltonians are constructed using Abelian hyperbolic band theory. In Secs.~\ref{sec:hyperAFM} and \ref{sec:hyperalt} we classify hyperbolic lattices into antiferromagnetic and altermagnetic categories based on their spin splitting and symmetries. In Sec.~\ref{sec:nonabelian}, we give a brief outlook beyond the Abelian sector using higher-genus supercells. We conclude in Sec.~\ref{sec:Conclusions} with the main findings of our work.

\section{Hyperbolic lattice model}
\label{sec:theory}

In order to give the reader a better understanding of our findings, we start with an introductory section explaining the hyperbolic space, Abelian band theory, and our model.

\subsection{Hyperbolic lattice and crystallography}
As mentioned above, regular hyperbolic lattices are described by their Schläfli symbol $\{p,q\}$, where $p$ denotes the number of edges of the polygon used as the building block and $q$ refers to the number of polygons meeting at each vertex. In flat Euclidean space, it can be shown geometrically that only the $\{4,4\}$, $\{3,6\}$, and $\{6,3\}$ lattices tile the plane without gaps, as they satisfy the equality $(p-2)(q-2)=4$. In negatively curved hyperbolic space, this requirement is relaxed to
\begin{equation}
    \label{eq:hyperrule}
    (p-2)(q-2)>4,
\end{equation}
yielding infinitely many possible regular lattice constructions. A convenient way to visualize hyperbolic lattices is by using the Poincaré disk model. In this representation, the entire infinite hyperbolic plane is mapped onto a unit disk. Starting from a single regular $p$-gon in the center, the $\{p,q\}$ lattice is constructed by repeatedly reflecting or copying that polygon across its edges such that $q$ polygons meet at every vertex, until a tessellation of the desired size is reached \cite{hypertiling}. 

\begin{figure}
    \centering
    \includegraphics[width=\linewidth]{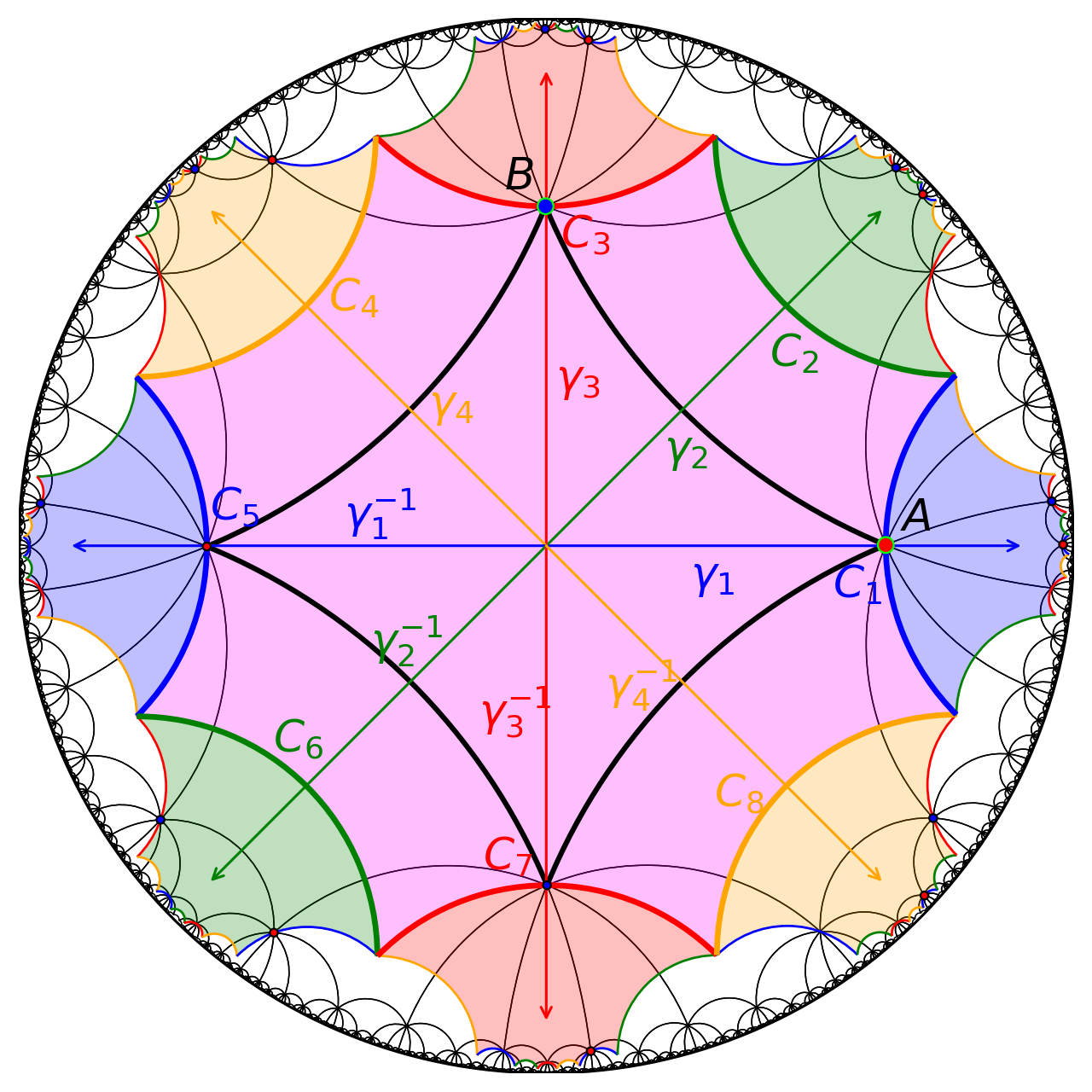}
    \caption{Example of the $\{4,8\}$ hyperbolic atomic lattice (black) made of eight squares that meet at each vertex. The colored octagons represent the central unit cell (pink) and its neighboring cells (blue, green, red, yellow). The similarly colored arrows $\gamma_i$ pointing towards the colored edges $C_i$ represent the four main translation directions of this lattice in which it is periodic. The $A$ and $B$ sublattices are shown in the central unit cell (green border) as red and blue dots respectively. Also shown are the $A$ and $B$ atoms at their correct locations in the neighboring colored cells.}
    \label{fig:example}
\end{figure}

The $\{4,8\}$ hyperbolic lattice is shown as an example in Fig.~\ref{fig:example}.
The resulting lattice can be seen as a graph whose vertices correspond to atomic sites and whose edges represent nearest-neighbor connections. As we are interested in the electronic structure of these lattices, we require a periodic description of the system, which allows us to probe momentum space via Fourier transformation. To this end, we work with a finite repeating patch that serves as a fundamental unit. We impose periodic boundary conditions by identifying pairs of boundary edges of the finite patch. This produces a compact closed surface of genus $g$, whose covering space is the full hyperbolic lattice. This construction is closely related to Euclidean Bloch theory, where the unit-cell structure can always be mapped onto a torus of genus $g=1$ in $2$D. In the case of the square lattice, one may think of gluing the opposing edges of the fundamental unit cell together in order to form such a torus. Using this construction yields a momentum vector with dimensionality $2g$. As hyperbolic lattices require at least $g=2$, the momentum space and the Brillouin zone (BZ) are at least $4$D. Note, that the real space lattice still remains two-dimensional. 

Abelian hyperbolic band theory then follows by introducing phase twists along the independent non-contractible cycles of the chosen unit cell. These twists play the role of generalized crystal momenta and allow one to define a Bloch-like Hamiltonian for a given hyperbolic lattice. In this way, the band structure is obtained from a finite unit cell together with the phases one collects when crossing boundaries into neighboring cells. This is, in principle, analogous to the way Euclidean band theory works, where one can think of collecting the phases acquired through the Fourier transformation of the periodic system when crossing into neighboring unit cells. For some lattices, the choice of the fundamental polygon is such, that even though the lattice possesses genus $g$, the fundamental polygon has more than $2g$ sides. In these cases, $2g$ of the phase twists are non-contractible, while the rest can be constructed from them.

In this work, we take the hyperbolic families introduced in Ref.~\cite{Boettcher2022Mar} and examine their electronic band structure using Abelian band theory in the context of altermagnetic order. The list of families is shown in Table~\ref{tab:families}.
\begin{table}[ht]
\centering
\caption{The infinite families of atomic hyperbolic lattices $\{p,q\}$ and their corresponding Bravais lattices $\{p_{B},q_{B}\}$ as a function of genus $g$. For each family, $N$ denotes the number of sites in the unit cell \cite{Boettcher2022Mar}.}
\begin{ruledtabular}
\label{tab:families}
\begin{tabular*}{\linewidth}{@{\extracolsep{\fill}}lcr@{}}
$\{p,q\}$ & $\{p_{B},q_{B}\}$ & $N$ \\
\hline
$\{4g,4g\}$ & $\{4g,4g\}$ & $1$ \\
$\{2g+1,\,2(2g+1)\}$ & $\{2(2g+1),\,2g+1\}$ & $1$ \\
$\{2(2g+1),\,2g+1\}$ & $\{2(2g+1),\,2g+1\}$ & $2$ \\
$\{4g,4\}$ & $\{4g,4g\}$ & $2g$ \\
$\{2(2g+1),\,3\}$ & $\{2(2g+1),\,2g+1\}$ & $2(2g+1)$ \\
\end{tabular*}
\end{ruledtabular}
\end{table}

As we aim to construct the lattice by placing spin-up and spin-down moments in an alternating fashion, we require the lattice to be bipartite. We therefore rule out the first two families with $N=1$. In addition to the families with even $N$, we also examine exceptional cases \cite{Boettcher2022Mar} that do not fit into the infinite families. To this end, we analyze the $\{4,8\}$ and $\{8,3\}$ lattices, both of which have an $\{8,8\}$ Bravais lattice.

\subsection{Tight-binding model}
Creating an altermagnetic model in Euclidean space usually consists of two parts. First, one has to find a magnetic ordering that leaves the total magnetization across the crystal compensated. Since we focus on collinear models, the easiest way to achieve this is by alternating the direction of the local atomic moment when moving across the lattice. The simplest Euclidean example of this model is a system with Néel order.
Such a system obeys an antiunitary symmetry operation connecting the two spin bands and real space sublattices. These are typically inversion combined with time reversal $\mathcal{PT}$ or real space translations combined with time reversal $\tau\mathcal{T}$. For this reason, the second step is introducing a mechanism which breaks these degeneracy protecting symmetries, while still preserving symmetry under rotation combined with time reversal. This can be achieved by introducing orbital ordering with next-nearest neighbor (NNN) hopping, 
by including non-magnetic sites in the unit cell \cite{Brekke2023Dec},
or by constructing spin clusters \cite{Zhu2025Apr}. 

In this work we assume the examined lattices to have a non-frustrated Néel magnetic ordering. While we do not employ a self-consistent derivation of this magnetic ordering, we note that it is not an arbitrary ansatz. Self-consistent antiferromagnetic solutions of the repulsive Hubbard model on several bipartite hyperbolic lattices have been identified \cite{Gluscevich2025Mar}. Additionally, another work studying Hubbard and Heisenberg models on certain bipartite hyperbolic lattices obtained results that are consistent with a robust antiferromagnetic ordering using Hartree-Fock mean-field, spin-wave and quantum Monte Carlo methods \cite{Gotz2024Dec}.

We keep the construction as simple as possible and just utilize the underlying hyperbolic point groups to create altermagnetic order. To this end, we focus on regular hyperbolic lattices with isotropic hopping up to NNN. Our spin-dependent Hamiltonians take the general form
\begin{align}
\begin{split}
    \label{eq:generalham}
    \mathcal{H}_s=&-t\sum_{ \langle i,j\rangle ,\sigma}c_{i,\sigma}^\dagger c_{j,\sigma} -t_2\sum_{\langle \langle i,j\rangle \rangle,\sigma}c_{i,\sigma}^\dagger c_{j,\sigma} \\
    &-J\sum_{i,\sigma,\sigma'}\mathbf{S}_i \cdot \boldsymbol{\sigma}_{\sigma,\sigma'} c_{i,\sigma}^\dagger c_{i,\sigma'},
\end{split}
\end{align}
where $\langle\rangle$ denotes nearest neighbor (NN) hopping and $\langle\langle\rangle\rangle$ NNN hopping. The operators $c_{i,\sigma}^\dagger$ and $c_{i,\sigma}$ create and annihilate an electron with spin $\sigma$ at lattice site $i$, respectively. 
The itinerant electrons move between (next-)nearest neighbors with hopping strength $t$ ($t_2$) and couple to the lattice local magnetic moment $\mathbf{S}_i$ at site $i$ with strength $J$.
The spin of the itinerant electrons at site $i$ is represented by the vector $\sum_{\sigma,\sigma'}\boldsymbol{\sigma}_{\sigma,\sigma'}c_{i,\sigma}^\dagger c_{i,\sigma'}$. We assume, that the local spins of the lattice point in the $z$ direction and absorb the magnitude of $S_i^z$ into $J$. This simplifies the coupling term to $\mp J\sum_{i,\sigma}\sigma c_{i,\sigma}^\dagger c_{i,\sigma}$ for the local spin-up and spin-down sites, respectively. 
We thus construct our model by placing local magnetic moments on a given hyperbolic lattice in an alternating fashion. For the cases we examine, this yields a spin compensated, unfrustrated unit cell and, by extension, a fully compensated crystal. We denote the atoms with a local spin-up moment with on-site energy $-J\sigma$ as $A$-type and spin-down sublattice with $+J\sigma$ as $B$-type. The momentum-space Hamiltonian emerging from this model is constructed by using the previously described hyperbolic band theory \cite{Maciejko2021Sep}.

Note, that any bipartite model that only uses NN hopping between the two types of sublattices is analytically degenerate as shown in Appendix~\ref{app:degeneracy}. Because of this, we omit NN hopping unless stated otherwise and focus only on NNN hopping for analytic convenience. In general we expect the NN hopping $t$ to be greater than the NNN hopping $t_2$.

\section{Spin degenerate lattices in the Abelian sector} \label{sec:hyperAFM}
\begin{figure*}[htb]
    \centering
    \includegraphics[width=0.95\linewidth]{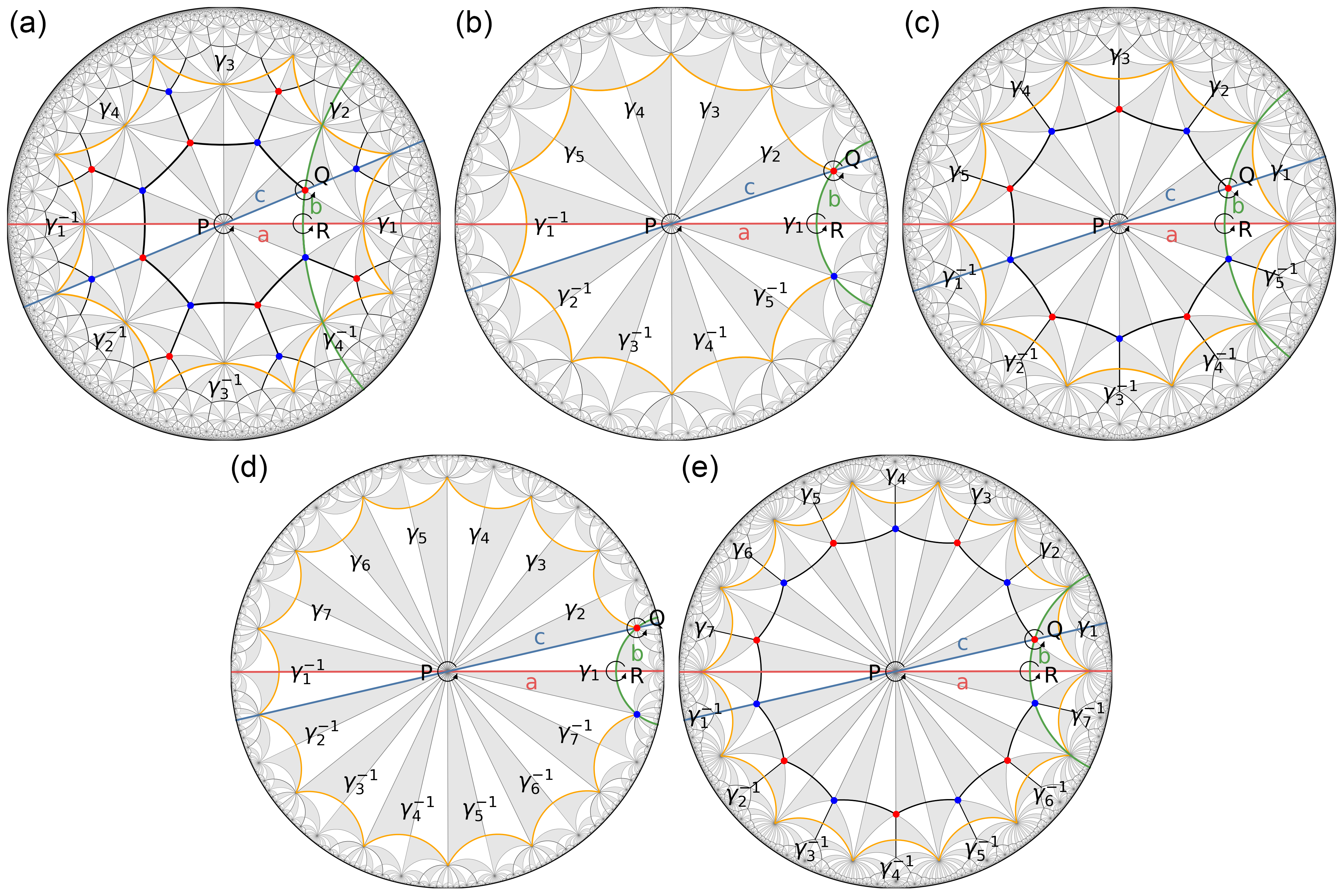}
    \caption{Examples of antiferromagnetic bipartite hyperbolic lattices (black) and their Schwarz triangles (gray/white). Atoms in the unit cell (orange) are colored red (spin-up) and blue (spin-down). The edges of the primitive unit cell are labeled with their generators of translations $\gamma_i$ in the corresponding directions. Also shown are the mirror operations a,b and c (red, green, blue) and the rotations P,Q and R. The first row shows $g=2$ lattices with the exceptional $\{8,3\}$ lattice (a), the $\{10,5\}$ lattice (b) and the $\{10,3\}$ lattice (c). The second row shows $g=3$ lattices $\{14,7\}$ (d) and $\{14,3\}$ (e). Note that (b) and (d) belong to the $\{2(2g+1),2g+1\}$ family while (c) and (e) belong to the $\{2(2g+1),3\}$ family.}
    \label{fig:antiferro}
\end{figure*}

Let us first discuss the lattices without spin-split bands in the Abelian sector. We find this to be the case for the $\{2(2g+1),2g+1\}$ and the $\{2(2g+1),3\}$ families of lattices. The exceptional case $\{8,3\}$ also shows no spin splitting. The lattices of genus $g=2,3$ for the two families and the $\{8,3\}$ lattice are shown in Fig.~\ref{fig:antiferro} with their corresponding geometric symmetry operations. Note, that for the $\{10,5\}$ and $\{10,3\}$ lattices, one of the generalized crystal momenta is reducible to $k_5=-(k_1-k_2+k_3-k_4)$. Similarly, one has $k_7=-(k_1-k_2+k_3-k_4+k_5-k_6)$ for the $\{14,7\}$ and $\{14,3\}$ lattices.
In the case of the $\{2(2g+1),\,2g+1\}$ family, the spin degeneracy within the Abelian hyperbolic band theory is enforced by $\mathcal{PT}$ symmetry, with the center of the unit cell as the inversion center. In real space this combination of operations acts as a sublattice exchange. In this case, inversion $\mathcal{P}$ is also equivalent to $P^{p/2}=P^{2g+1}$ in terms of hyperbolic point group operations. Global spin degeneracy therefore holds for the entire family within the corresponding Abelian Brillouin zone. This statement does not, by itself, determine the behavior of every higher-dimensional representation sector.

The same argument can be made for the $\{2(2g+1),\,3\}$ family of lattices. Here, $\mathcal{PT}$ around the unit cell center also acts as a sublattice exchange in real space. Inversion can once again be represented by $P^{p/2}=P^{2g+1}$. Consequently, these lattices remain spin degenerate across the Abelian Brillouin zone.
For the $\{8,3\}$ lattice, $P^4$ acts as inversion. In this case, however, it does not correspond to a sublattice exchange symmetry when combined with time reversal. Sublattice exchange operations like bond inversions $R\mathcal{T}$ or translations $P^4R\mathcal{T}$ do not map $\vec{k}\to \vec{k}$ and thus only enforce degeneracies in certain subspaces. For this lattice, we find analytically that $t_A(\vec{k})$ and $t_B(\vec{k})$ yield the same characteristic polynomial. For this reason, the spin-up and spin-down energy spectra have to be equivalent. This case is an example of a lattice, where total spin degeneracy is not enforced by symmetry, but where it does exist in practice. This shows that, while spin splitting is allowed in the system, it does not necessarily need to exist. The NNN hopping TB Hamiltonians for the $g=2$ cases and their relevant point group operations can be found in Appendix~\ref{app:anti}.

Although these lattices are spin degenerate in the primitive-cell Abelian sector, this does not by itself determine the spectra of all higher-dimensional translation representations. A supercell calculation \cite{Lenggenhager2023Dec} can access non-Abelian eigenstates beyond the Abelian theory. In Sec.~\ref{sec:nonabelian}, we show that the $\{10,5\}$ lattice provides an example in which two such supercells exhibit spin splitting, despite the degeneracy of the Abelian spectrum.

\section{Hyperbolic altermagnets}
\label{sec:hyperalt}
In this section, we show that the examined members of the $\{4g,4\}$ family of hyperbolic lattices and the exceptional $\{4,8\}$ lattice exhibit altermagnetic order within our simple model when analyzed using Abelian hyperbolic band theory. In contrast to the previous section, the presence of a band splitting in the Abelian sector is sufficient to classify the lattice as an altermagnet, without an analysis of the non-Abelian states. This is because Abelian states remain valid in the full, non-Abelian band theory. Consequently, the possible additional degenerate bands in the full non-Abelian spectrum do not affect this classification. We use Fermi surface cuts at constant energies as one possible way to examine the energy splitting. Since the momentum space in hyperbolic band theory has at least four dimensions, we cut two-dimensional slices out of it to visualize the electronic band structure. We vary two momentum components $k_i,k_j$ while setting all other components to $k_{k}=0$. These are the simplest possible $2$D cuts. We want to stress, that these cuts are primarily used as a means of visualizing the band structure of the $4$D momentum space. While we also identify certain known Euclidean splittings in these plots, they are meant to be seen as remarks on the splitting in these subspaces. They should not be taken as classifications of the full momentum space. For these spin-split cases, we also include NN hopping for completeness but again note, that it is not the origin of the splitting. 

\subsection{The \texorpdfstring{$\{4g,4\}$}{\{4g,4\}} family}
\begin{figure*}
    \centering
    \includegraphics[width=0.95\linewidth]{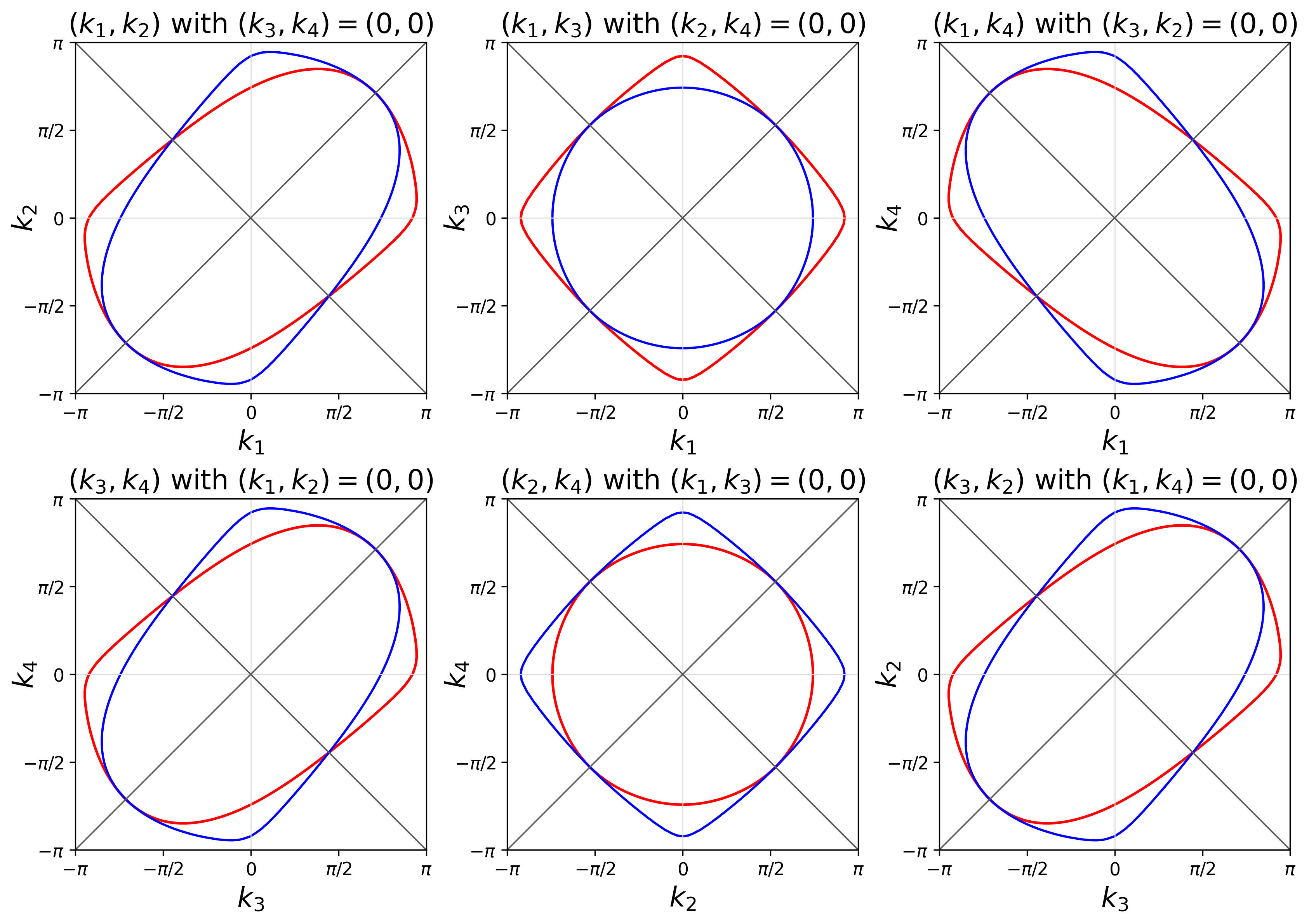}
    \caption{$2$D constant energy Fermi surface cuts of the Abelian $4$D momentum space of the $\{8,4\}$ hyperbolic lattice with spin-up bands shown in red and spin-down bands colored blue. We observe a general spin splitting with degeneracies only enforced on certain nodal lines. These are given by the symmetry operations shown in Table~\ref{tab:848sym}. In the $(k_1,k_2/k_4)$ and $(k_3,k_2/k_4)$, subspaces we observe a deformed $d$-wave like altermagnetic pattern with spin flavors switching between the two nodal lines. In $(k_1,k_3)$ and $(k_2,k_4)$, the spectrum takes the shape of an anisotropic $s$-wave, where we still retain two nodal lines but observe no spin switching between the degeneracies. In all plots, we set $t=t_2=J/2=-\mu/7$.}
    \label{fig:84cuts}
\end{figure*}

We explicitly examine the $\{8,4\}$ lattice shown in Fig.~\ref{fig:84} as the lowest genus example with $g=2$. The lattice Hamiltonian is included in Appendix \ref{app:84}. We plot the six simplest $2$D Fermi surface cuts in Fig.~\ref{fig:84cuts} illustrating spin splitting and nodal lines.
\begin{figure*}
    \centering
    \includegraphics[width=0.95\linewidth]{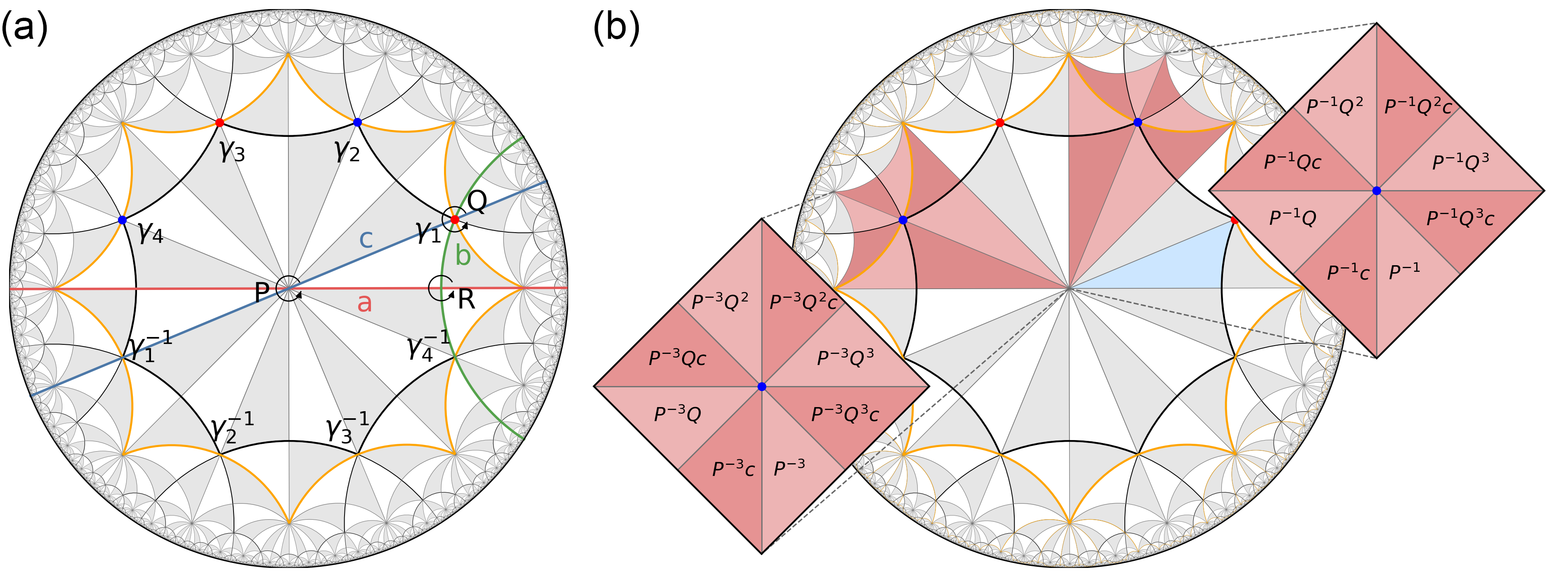}
    \caption{(a) Bipartite $\{8,4\}$ lattice (black) as the lowest genus ($g=2$) example of the $\{4g,4\}$ family of lattices with its Schwarz triangles (gray/white). The four atoms in the central primitive unit cell (orange) are colored red (spin-up) and blue (spin-down). Labels $\gamma_i$ denote the generators of translations while a,b and c (red, green, blue) mark the fundamental mirrors of the lattice. P, R and Q represent the rotation elements of the point group. (b) The 16 degeneracy-enforcing symmetries of the form $\mathcal{E}_{\{8,4\}}=\{ P^{-i}Q^jc^k:i \in \{1,3\},\;j \in \{0,\dots,3\},\;k \in \{0,1\}\}$, which map a Schwarz triangle from the $A$ sublattice (blue) to one of the triangles of the $B$ sublattice (red).}
    \label{fig:84}
\end{figure*}
The resulting Fermi surface cuts are highly unconventional. In the $(k_1,k_2/k_4)$ and $(k_3,k_2/k_4)$ subspaces the spectra resemble a deformed $d$-wave splitting with four nodal lines along points of equivalent momenta. The remaining  $(k_1,k_3)$ and $(k_2,k_4)$ subspaces instead yield an anisotropic $s$-wave splitting which also hosts four similar nodal lines but does not show a sign-change between them. 

The reason why this lattice exhibits spin-dependent splitting can be explained by its symmetries. Hyperbolic lattices can be described by hyperbolic point groups denoted by \textbf{P}$(2,p,q)$ \cite{Chen2023Aug}, so this system has the point group \textbf{P}$(2,4,8)$. In principle, this means that $P^4$ acts as inversion with the unit cell center as the inversion center. The crucial difference to the previously described antiferromagnetic cases, is that this inversion is not an operation that exchanges sublattice types and thus does not leave the system invariant when combined with time reversal symmetry $\mathcal{T}$. The system also does not possess any translation operation that exchanges sublattices and maps $\vec{k}\to \vec{k}$ in combination with time reversal symmetry. It therefore lacks a symmetry operation that enforces $\operatorname{spec}\bigl(H_A(\vec{k},\sigma)\bigr)=\operatorname{spec}\bigl(H_B(\vec{k},-\sigma)\bigr)$ for a generic $\vec{k}$ and consequently allows for a band-splitting between spin-up and spin-down bands to exist. Note, that `$\operatorname{spec}$' denotes the eigenvalue spectrum of the Hamiltonian.
This model of a compensated real space lattice with spin-split electron bands thus qualifies as an altermagnet by definition. 

The splitting given by the two bands of lowest energy assuming only NNN can be written in the compact form
\begin{widetext}
\begin{align}
    \Delta_{\{8,4\}}
    &= 2 t_2 \Biggl(
    \cos\left(\frac{1}{2}(k_1-2k_2+k_3)\right)
    - \cos\left(\frac{1}{2}(2k_1-k_2+k_4)\right)
    + \cos\left(\frac{k_1-k_3}{2}+k_4\right) \notag\\
    &\quad
    + 2 \cos\left(\frac{k_1}{2}\right)\cos\left(\frac{k_3}{2}\right)
    - \cos\left(\frac{1}{2}(k_2-2k_3+k_4)\right)
    - 2 \cos\left(\frac{k_2}{2}\right)\cos\left(\frac{k_4}{2}\right)
    \Biggr).
\end{align}
\end{widetext}
Note that this splitting does not contain $J$. This is because when we set $t=0$, both sublattice blocks effectively decouple. As a consequence, $J$ cancels when taking the difference between the eigenvalues of both blocks. The result represents an even-wave splitting as expected. In Euclidean space, altermagnetic patterns can be classified by their symmetries in spin and real space given by the notation $[R_i||R_j]$, where $R_i$ is a transformation acting on spin space, and $R_j$ is a transformation acting on real space. For example, a $g$-wave altermagnet is defined by its eightfold rotational symmetry in real space combined with a twofold spin rotation, yielding $[C_2||C_8]$. This symmetry is also reflected in the electronic band structure, where $d$-, $g$-, $i$-, and higher order splittings may be observed. These splittings are named after their resemblance to the angular structures of the corresponding atomic orbitals. Hence, the type of altermagnetic symmetry present can also be classified by examining the electron bands in momentum space.

Since the $k$-space in our case is four-dimensional, we instead use generalized classifications of $n$-dimensional orbitals \cite{Hosoya1995Jun}. We note, that this classification is purely mathematical. We expand the splitting in $\vec{k}$ up to the lowest non-vanishing order, yielding 
\begin{widetext}
\begin{equation}
    \Delta_{\{8,4\}}=-\frac{t_2}{16}(k_1-k_2+k_3-k_4)(k_1-k_2-k_3+k_4)(k_1+k_2-k_3+k_4)(k_1-k_2+k_3+k_4)+\mathcal{O}(k^6).
\end{equation}
\end{widetext}
In this expansion, the zeroth order and the $k^2$ polynomial cancel exactly. We refer to the remaining order $k^4$ polynomial as $P_{\{8,4\}}$. This expression does not satisfy $\Delta_4P_{\{8,4\}}=\sum_{i=1}^4\frac{\partial^2}{\partial k_i^2}P_{\{8,4\}}=0$, and is therefore not a harmonic polynomial. It can be decomposed into
\begin{equation}
    P_{\{8,4\}}=t_2\left( P_g + |\vec{k}|^2P_d \right),
\end{equation}
with $P_d=\frac{1}{32}(-k_1^2+k_2^2-k_3^2+k_4^2)$ being a superposition of $4$D $d$-wave orbitals of order $k^2$ and satisfying $\Delta_4P_d=0$. Because they are  multiplied with a radial scalar $|\vec{k}|^2$, they still appear at order $k^4$. However, the radial component $|\vec{k}|^2$ does not add any new nodes, besides the trivial node at $\vec{k}=0$. The remainder is a $g$-wave orbital construction of order $k^4$ that satisfies $\Delta_4P_g=0$. Both $P_g$ and $P_d$ can be constructed by $4$D $d$- and $g$-wave orbitals found in Ref.~\cite{Hosoya1995Jun}. The system therefore possesses $g$- and $d$-wave character. As both parts scale linearly with $t_2$, none of the two is predominant over the other as long as $t=0$. Once NN hopping is included, the parameters $J,t,\mu$ and the scale of the Fermi wave-vector $k_F$, at which the splitting is examined, may lead to one of the two orbital characters outweighing the other. We conjecture that both $d$-wave and $g$-wave altermagnetism coexist on this hyperbolic lattice.
Just like in Euclidean space, we find spin degenerate subspaces in the, in our case, four-dimensional momentum space. In certain subspaces, spin degeneracies are enforced by the point group elements of \textbf{P}$(2,4,8)$ \cite{Chen2023Aug} that are still valid after making the lattice bipartite. The point group possesses 32 distinct elements and the matrices of the six symmetry generators are listed in Appendix \ref{app:84}. We search the point group for operations that exchange the $A$-type and $B$-type sublattices and yield $t_A(\Omega \vec{k}) \leftrightarrow t_B(\vec{k})$ at a spectral level, where $\Omega$ is a momentum transformation constituted by the generators of the symmetries combined with time reversal $\mathcal{T}$. We find the 16-element coset of the form $\mathcal{E}_{\{8,4\}}=\{ P^{-i}Q^jc^k:i \in \{1,3\},\;j \in \{0,\dots,3\},\;k \in \{0,1\}\}$. These elements can be thought of as the operations that map a Schwarz triangle adjacent to an $A$ atom to all 16 possible $B$ adjacent Schwarz triangles as seen in Fig.~\ref{fig:84}. 

Just like in the Euclidean altermagnetic case, the degeneracy-preserving symmetry operations contain rotations around the unit cell center. Here they take the form $P^{-1}$ and $P^{-3}$. The bond-center rotation $R=P^{-1}Q^3$ is also a valid symmetry, and so are the rotations $PRP^{-1}=P^{-3}Q^3,P^2RP^{-2}=P^{-1}Q$ and $P^3RP^{-3}=P^{-3}Q$ around the other bond centers. In addition to the aforementioned rotations, we observe the simple bond mirrors $a=P^{-1}c,PaP^{-1}=P^{-3}Q^2c,P^2aP^{-2}=P^{-1}Q^2c$ and $P^3aP^{-3}=P^{-3}c$ enforcing spin degeneracies. These bond symmetries are usually broken in the Euclidean altermagnets. The remaining four symmetries are $P^{-1}Qc,P^{-1}Q^3c,P^{-3}Qc$ and $P^{-3}Q^3c$. 
Going back to Fig.~\ref{fig:84cuts}, it is possible to identify the symmetries that yield certain nodal lines. This is shown in Table~\ref{tab:848sym}.
\begin{table}[ht]
\centering
\caption{List of symmetry operations for the $\{8,4\}$ and $\{4,8\}$ lattice which produce nodal lines in Fig.~\ref{fig:84cuts} and Fig.~\ref{fig:48cuts}. Visual interpretations of these elements can be seen in Fig.~\ref{fig:84} and Fig.~\ref{fig:48excep}.}
\label{tab:848sym}
\begin{ruledtabular}
\begin{tabular}{l@{\hspace{1.5em}}l@{\hspace{1.5em}}l@{\hspace{1.5em}}l}
Subspace & Nodal line & $\Omega_{\{8,4\}}$ & $\Omega_{\{4,8\}}$ \\
\hline
$(k_1,k_2)$ & $k_1=k_2$   & $P^{-3}Q^2c,\;P^{-3}Q$ & \\
$(k_1,k_2)$ & $k_1=-k_2$  & $P^{-3}c$ & \\
$(k_1,k_2)$ & $k_1=0$     & & $P^{-1}Q^7,\;P^{-1}Q^4c$ \\
\hline
$(k_1,k_3)$ & $k_1=k_3$   & & $P^{-1}Q^5,\;P^{-1}Q^4c$ \\
$(k_1,k_3)$ & $k_1=-k_3$  & & $P^{-1}c,\;P^{-1}Q^7$ \\
\hline
$(k_1,k_4)$ & $k_1=k_4$   & $P^{-1}Q^2c$ & \\
$(k_1,k_4)$ & $k_1=-k_4$  & $P^{-1}c,\;P^{-1}Q$ & \\
$(k_1,k_4)$ & $k_1=0$     & & $P^{-1}c,\;P^{-1}Q^5$ \\
\hline
$(k_2,k_3)$ & $k_2=k_3$   & $P^{-1}Q^3,\;P^{-1}Q^2c$ & \\
$(k_2,k_3)$ & $k_2=-k_3$  & $P^{-1}c$ & \\
$(k_2,k_3)$ & $k_3=0$     & & $P^{-1}Q^7,\;P^{-1}Q^4c$ \\
\hline
$(k_2,k_4)$ & $k_2=0$     & & $P^{-1}c,\;P^{-1}Q^5$ \\
$(k_2,k_4)$ & $k_4=0$     & & $P^{-1}Q^7,\;P^{-1}Q^4c$ \\
\hline
$(k_3,k_4)$ & $k_3=k_4$   & $P^{-3}Q^3,\;P^{-3}c$ & \\
$(k_3,k_4)$ & $k_3=-k_4$  & $P^{-3}Q^2c$ & \\
$(k_3,k_4)$ & $k_3=0$     & & $P^{-1}c,\;P^{-1}Q^5$ \\
\end{tabular}
\end{ruledtabular}
\end{table}
We find that the nodal lines in the  $(k_1,k_3)$ and $(k_2,k_4)$ subspaces are not enforced by an underlying symmetry, but instead are due to these specific choices of cuts and thus accidental nodal lines.

For higher-genus lattices of this family, we also expect spin splitting to be present, since the inversion operation $\mathcal{P}=P^{2g}$ never exchanges atom types for any member of the family. Therefore, there exists no symmetry which leaves the real space lattice invariant in combination with $\mathcal{T}$ and enforces $\operatorname{spec}\bigl(H_A(\vec{k},\sigma)\bigr)=\operatorname{spec}\bigl(H_B(\vec{k},-\sigma)\bigr)$ at arbitrary $\vec{k}$. Hence, spin splitting is allowed, except in the degenerate subspaces enforced by the surviving point groups. Accidental degeneracies that leave the spectrum without spin splitting, however, cannot be ruled out. We have explicitly tested this for the $\{12,4\}$ lattice of genus $g=3$ (not included in this work) and could again identify spin splitting with certain degeneracies being protected by the symmetries of the \textbf{P}$(2,4,12)$ point group.  

\subsection{The \texorpdfstring{$\{4,8\}$}{\{4,8\}} exceptional lattice}

\begin{figure*}
    \centering
    \includegraphics[width=0.95\linewidth]{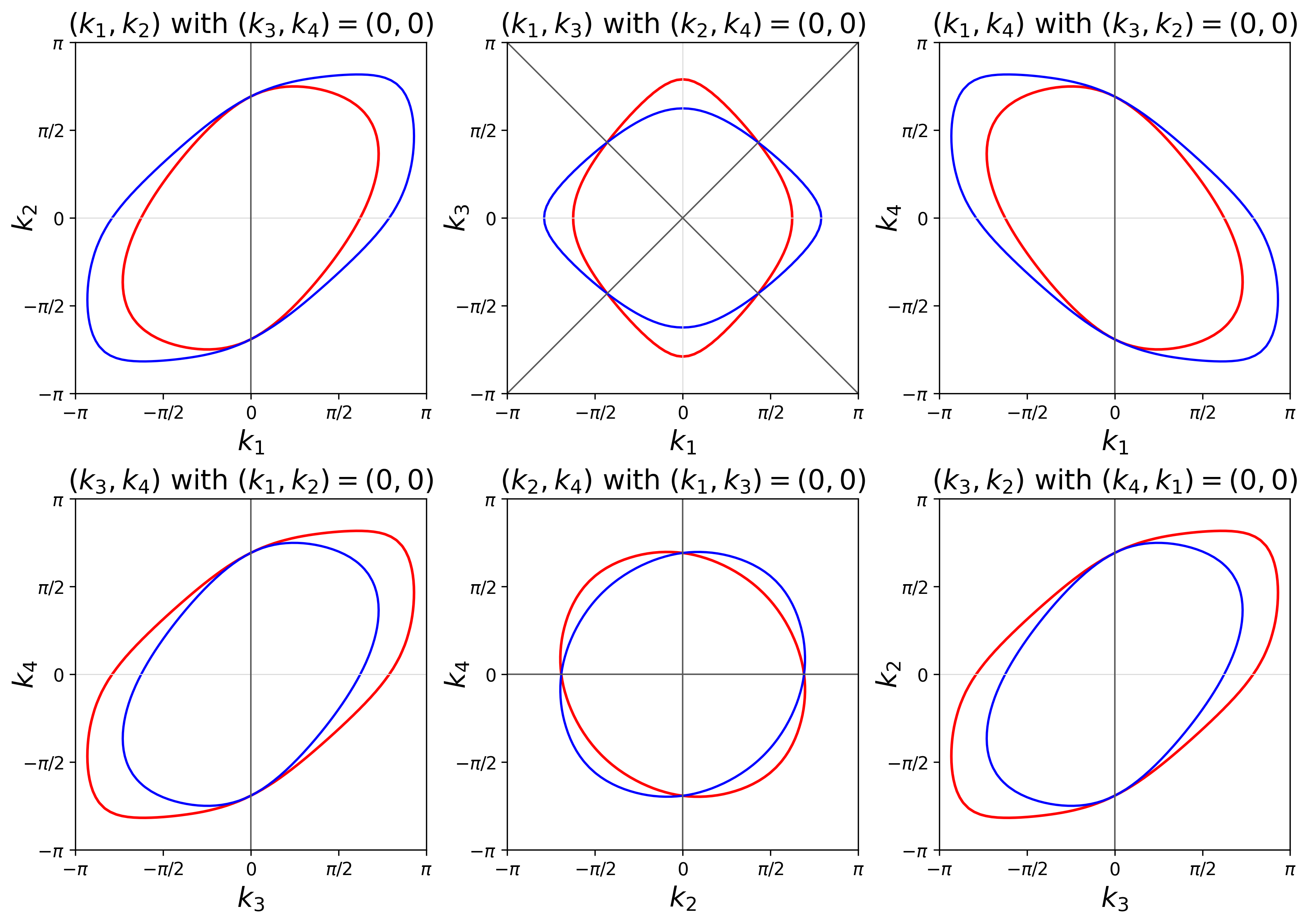}
    \caption{$2$D constant energy Fermi surface cuts of the Abelian $4$D momentum space of the $\{4,8\}$ hyperbolic lattice with spin-up bands shown in red and spin-down bands colored blue. We observe spin splitting with degeneracies only enforced on certain nodal lines. These are again given by the symmetry operations shown in Table~\ref{tab:848sym}. In the $(k_1,k_2/k_4)$ and $(k_3,k_2/k_4)$ subspaces, we observe only a single line crossing into the subspace. In between, the spin flavor does not switch. In $(k_1,k_3)$ and $(k_2,k_4)$, the spectrum takes the exact shape of a $d_{x^2-y^2}$ and $d_{xy}$ altermagnet known from Euclidean space with four nodal lines and spins switching between them. In all plots, we set $t=t_2=J/2=-\mu/7$.}
    \label{fig:48cuts}
\end{figure*}

We repeat the same process as for the $\{8,4\}$ lattice with the $\{4,8\}$ exceptional lattice shown in Fig.~\ref{fig:48excep} and its lattice Hamiltonian included in Appendix~\ref{app:48}. Plots of the six simple $2$D Fermi surface cuts are depicted in Fig.~\ref{fig:48cuts}.
\begin{figure*}
    \centering
    \includegraphics[width=0.95\linewidth]{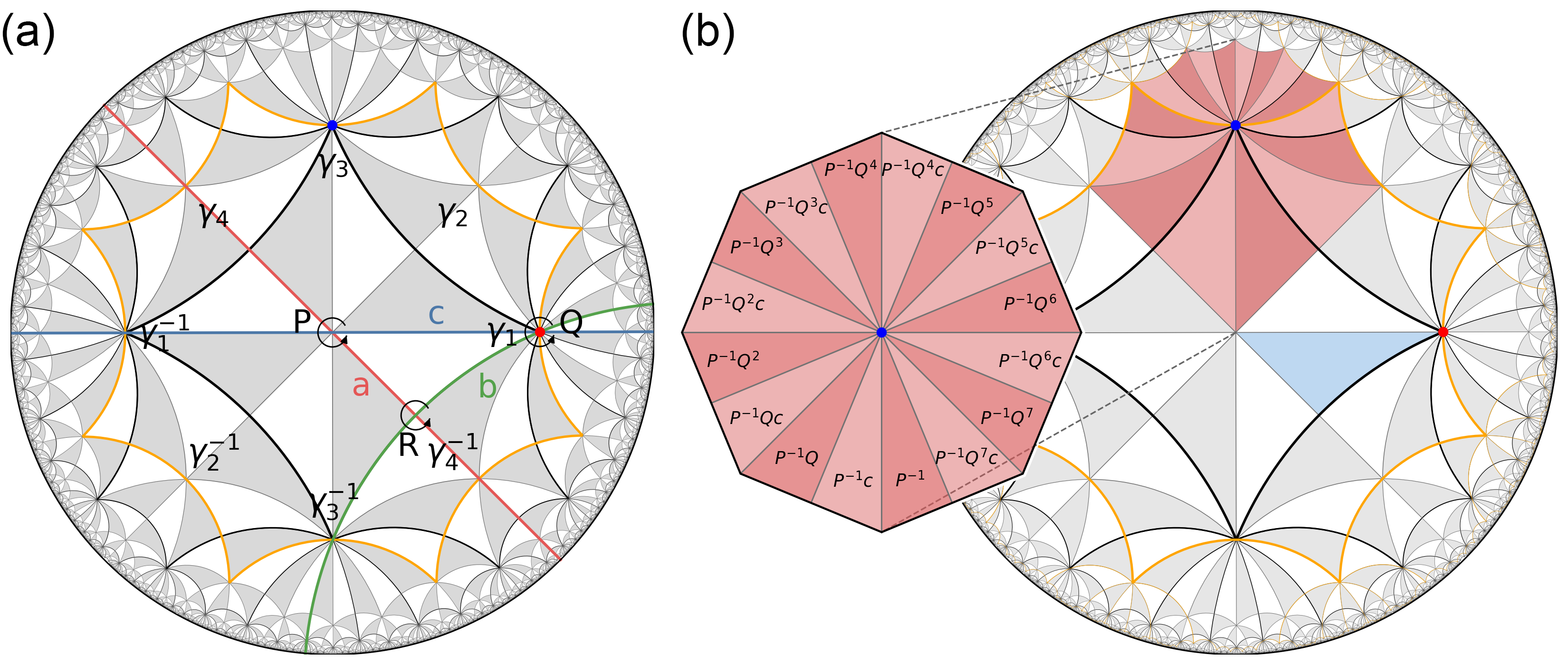}
    \caption{(a) Bipartite exceptional $\{4,8\}$ lattice (black) with its Schwarz triangles (gray/white). This lattice contains two atoms in its primitive unit cell (orange), each hosting a spin-up moment (red) and a spin-down (blue) respectively. $\gamma_i$ are the generators of translation in the shown directions and a,b and c (red, green and blue) depict the basic mirrors. The generators of rotation are labeled P, Q and R. (b) The 16 degeneracy-enforcing symmetry operations $\mathcal{E}_{\{4,8\}}=\{P^{-1}Q^jc^k: j\in\{0,\dots 7\},k\in\{0,1\}\}$ which map a Schwarz triangle adjacent to atom $A$ (blue) to all possible triangles adjacent to atom $B$ (red).}
    \label{fig:48excep}
\end{figure*}
In the $(k_1,k_2/k_4)$ and $(k_3,k_2/k_4)$ subspaces, we only observe a single degenerate nodal line that does not flip spins when crossed. More interesting in this case are the $(k_1,k_3)$ and $(k_2,k_4)$ subspaces. Here, the spectra represent non-deformed $d$-wave altermagnetic patterns exactly. We specifically find a $d_{x^2-y^2}$-like pattern in the $(k_1,k_3)$ subspace and a $d_{xy}$-like pattern in the $(k_2,k_4)$ subspace. The full energy splitting between the lowest spin-up and spin-down electron bands takes the form 
\begin{equation}
\begin{split}
    \Delta_{\{4,8\}} = 2t_2 \Bigl(
    &C_{1}+C_{2\bar{3}4}+C_{1\bar{2}}+C_{14} \\
    &- C_{3}-C_{41\bar{2}}-C_{3\bar{4}}-C_{2\bar{3}}
    \Bigr)
\end{split}
\end{equation}
where we use the $C_{ijk\dots}=\cos\left( k_i+k_j+k_k+\dots \right)$ notation. For this splitting, we again use $4$D atomic orbitals for classification instead of only relying on the central rotational symmetry. Expanding in $k$ yields
\begin{equation}
    \Delta_{\{4,8\}} = -2t_2(k_1^2-k_3^2+2k_2k_4) + \mathcal{O}(k^4).
\end{equation}
Here, only the zeroth order cancels and we refer to the non-vanishing polynomial of order $k^2$ as $P_{\{4,8\}}$. In contrast to the $\{8,4\}$ lattice, $P_{\{4,8\}}$ satisfies $\Delta_4P_{\{4,8\}}=0$ by itself and needs no orbital decomposition. It is therefore a pure $4$D $d$-wave, and we find that $P_{\{4,8\}}$ can be constructed by $4$D $d$-wave orbitals found in \cite{Hosoya1995Jun}. Thus, the $\{4,8\}$ lattice is an example of a hyperbolic $d$-wave altermagnet. Since we consider a lattice model, there are also higher order terms which represent subdominant $g,i,\dots$-wave spin splittings.

This lattice does not fit into any of the infinite families listed in Table~\ref{tab:families} but it does have the same point group as the $\{8,4\}$ lattice. This is because the $\{8,4\}$ lattice is the dual lattice of the $\{4,8\}$ lattice. The only necessary step is to transform the elements into our chosen representation. The matrices of the symmetry generators are listed in Appendix~\ref{app:48}. Similarly as before, the 32 symmetry elements get reduced to 16 of the form $\mathcal{E}_{\{4,8\}}=\{P^{-1}Q^jc^k: j\in\{0,\dots 7\},k\in\{0,1\}\}$. These operations are again maps between the Schwarz triangles of both sublattices seen in Fig.~\ref{fig:48excep}. Crucially, this lattice also has no inversion elements $\mathcal{P}$, which exchanges both sublattices and thus enforce spin degeneracy across the full BZ through $\mathcal{P}\mathcal{T}$ symmetry. Examples of degeneracy-enforcing symmetries are the bond mirrors $a=P^{-1}c$ and $P^{-1}aP=P^{-1}Q^4c$, the rotation around the unit cell center $P^{-1}$ and the bond rotations $R=P^{-1}Q^7$ and $P^{-1}RP=P^{-1}Q^5$. We identify the symmetry elements that enforce nodal lines in Fig.~\ref{fig:48cuts} and list them in Table.~\ref{tab:848sym}. In this case the origins of all nodal lines seen in the Fermi surface cuts can be explained by the symmetries in $\mathcal{E}_{\{4,8\}}$.

\section{Outlook on Non-Abelian eigenstates}
\label{sec:nonabelian}

The results presented above are obtained within a primitive-cell Abelian hyperbolic band theory, in which the translation generators are represented by commuting phase factors. This sector is sufficient to establish altermagnetic behavior whenever spin splitting is present. The full hyperbolic translation group is, however, non-Abelian. Higher-dimensional representation sectors of the primitive translation group can be accessed by applying Abelian hyperbolic band theory to larger supercells \cite{Lenggenhager2023Dec}. In this section, we use this construction as a preliminary numerical outlook rather than a complete, non-Abelian symmetry analysis.

We focus on the $\{10,5\}$ lattice, whose Abelian spectrum is spin degenerate according to the symmetry argument discussed in Sec.~\ref{sec:hyperAFM}. Starting from its primitive genus-$2$ cell, we construct the higher-genus supercells of genus $6$ and $26$ using GAP \cite{GAP4}, HyperCells \cite{HyperCells} and HyperBloch \cite{HyperBloch}. The primitive cell contains two sites, while the genus-$6$ and genus-$26$ supercells contain $10$ and $50$ sites, respectively. In all three cases, the lattice remains bipartite and contains equal numbers of spin-up and spin-down sites.

For each cell we calculate
\begin{equation}
    D_{\mathrm{RMS}}=\sqrt{\frac{1}{N_kN}\sum_{a=1}^{N_k}\sum_{n=1}^{N}\left[E_{\uparrow,n}(\vec{k}_a)-E_{\downarrow,n}(\vec{k}_a)\right]^2}
\end{equation}
as the root-mean-square splitting, where $N_k$ and $N$ denote the number of sampled $k$-points and bands respectively. Additionally, we define
\begin{equation}
    D_{\mathrm{max}}=\max_{a,n}\left| E_{\uparrow,n}(\vec{k}_a)-E_{\downarrow,n}(\vec{k}_a) \right|
\end{equation}
as the maximum splitting observed in the sampled set of $k$-space points.

For each cell, we sample $4\times10^3$ momentum points throughout the corresponding $2g$-dimensional Brillouin zone. The results are shown in Table~\ref{tab:nonabelian}.
\begin{table}[ht]
\centering
\caption{Numerical spin splitting analysis for the primitive cell and two supercells of the $\{10,5\}$ lattice. We use $t=t_2=1=J/2$. As a reference, setting $J=0$ or $t_2=0$ yields a numerical precision floor of $D_{\mathrm{RMS}}=10^{-15}$.}
\label{tab:nonabelian}
\begin{ruledtabular}
\begin{tabular}{l@{\hspace{1.5em}}l@{\hspace{1.5em}}l@{\hspace{1.5em}}l}
Genus & Sites/cell & $D_{\mathrm{max}}$ & $D_{\mathrm{RMS}}$\\
\hline
2 & 2 & $1.42\times10^{-14}$ & $2.45\times10^{-15}$  \\
6 & 10 & $3.00$ & $6.05\times10^{-1}$  \\
26 & 50 & $0.88$ & $1.63\times10^{-1}$ \\
\end{tabular}
\end{ruledtabular}
\end{table}
As expected, the genus-$2$ spectrum remains spin degenerate to numerical precision. This confirms our symmetry argument in Sec.~\ref{sec:hyperAFM} for the $\{10,5\}$ lattice. In contrast, both higher-genus supercells exhibit a non-negligible spin splitting at the sampled momenta.

This result does not invalidate the primitive-cell $\mathcal{PT}$ argument. That argument is formulated within the Abelian representations of the primitive translation group and enforces degeneracy of the two spin flavors at Abelian momenta. An Abelian representation of the supercell translation group generally corresponds to a higher-dimensional representation of the primitive translation group. In this framework, $\mathcal{PT}$ may not map an arbitrary supercell momentum back to itself. Consequently, spin degeneracy at the same momentum is not automatically enforced.

For comparison, we also examine the band structure of the $\{8,4\}$, $\{8,3\}$ and $\{10,3\}$ lattices using supercells. The corresponding genera and the number of sites of the supercell are listed in Tab.~\ref{tab:nonab2}.
\begin{table}[b]
\centering
\caption{Results for selected supercells of three additional lattices. A checkmark denotes spin splitting, while a cross denotes spin degeneracy to numerical precision in the tested supercell.}
\label{tab:nonab2}
\begin{ruledtabular}
\begin{tabular}{l@{\hspace{1.5em}}l@{\hspace{1.5em}}l@{\hspace{1.5em}}l}
Lattice & Genus & Sites & Splitting \\
\hline
$\{8,4\}$ & 3 & 8 & \checkmark\\
$\{8,3\}$ & 5 & 64 & $\times$ \\
$\{10,3\}$ & 5 & 40 & $\times$ \\
\end{tabular}
\end{ruledtabular}
\end{table}
Using the previously described momentum sampling we find the bands of the $\{8,3\}$ and $\{10,3\}$ to be degenerate, just like in the Abelian sector. In the same vein, the $\{8,4\}$ lattice exhibits spin splitting also when a larger unit cell is considered, as expected.

Note, that since a finite number of supercells accesses only selected higher-dimensional representation sectors, these numerical results should not be interpreted as a complete classification of their non-Abelian spectra. The supercell calculations therefore provide evidence, that a spin degenerate Abelian spectrum does not necessarily imply spin degeneracy in the non-Abelian sector. A full treatment would require an analysis of the irreducible representations of the primitive translation group and the action of the point-group symmetries on them similar to the analysis in the Abelian sector.

\section{Conclusions}
\label{sec:Conclusions}

In this article, we extend the known Euclidean concept of altermagnetism to the Abelian sector of hyperbolic band theory. Unlike the Euclidean case, we identify altermagnets in hyperbolic lattices without including any special orbital ordering or lattice manipulations. Instead we find altermagnetism to be an inherent feature of certain hyperbolic lattices.
We give a brief introduction to hyperbolic crystallography and Abelian band theory. We show that the inclusion of next-nearest neighbor hopping is sufficient to turn certain bipartite hyperbolic lattices into altermagnets, while others remain antiferromagnetic. We find no spin splitting in the $\{2(2g+1),2g+1\}$ and $\{2(2g+1),3\}$ families of lattices nor in the exceptional case $\{8,3\}$. In the case of the aforementioned infinite families, this can be explained by $\mathcal{P}\mathcal{T}$ symmetry enforcing spin degeneracy within the Abelian sector.

In contrast, we find that the infinite family $\{4g,4\}$ and the exceptional case $\{4,8\}$ lack this symmetry entirely, allowing for spin-split electronic band structures. In both cases, we examine the elements of the point groups in the Abelian framework that lead to spin degeneracies in certain momentum subspaces and find a coset of 16 elements. We utilize $2$D Fermi surface cuts of constant energies in order to visualize the $4$D momentum space for the $g=2$ cases. Using the degeneracy-enforcing elements of the point groups, we predict the emergence of nodal lines in these slices. In the $\{4,8\}$ lattice, we find that the spin splitting combined with the nodal lines yield altermagnetic $d_{x^2-y^2}$- and $d_{xy}$-like spectra known from Euclidean space. Using higher-dimensional classifications of atomic orbitals, we argue that the $\{8,4\}$ lattice possesses both $d$-wave and $g$-wave character at lowest order in $k$, while the $\{4,8\}$ lattice represents a $d$-wave hyperbolic altermagnet. 

In the selected higher-genus supercells relvant to non-Abelian hyperbolic band theory, the $\{10,5\}$ lattice exhibits a spin split spectrum. No splitting was found for the tested $\{8,3\}$ and $\{10,3\}$ supercells, while the tested $\{8,4\}$ supercell remains spin split.

Our findings provide a basis for future studies of phenomena related to altermagnetism in non-Euclidean geometries, for instance, magnonic properties. They could be experimentally realized by extending existing hyperbolic electrical or photonic lattices \cite{Kollar2019Jul,Zhang2022May,Huang2024Feb} to include a two-level system on each site, realizing the necessary pseudospin degree of freedom. In this context, magnetic metamaterials  offer a complementary design principle for constructing these artificial magnetic properties \cite{Cong2018Oct,Tuz2020Apr}. 

\begin{acknowledgments}
This work was supported by the Deutsche Forschungsgemeinschaft (DFG, German Research Foundation) project SFB 1170 and DFG through the Würzburg-Dresden Cluster of Excellence ctd.qmat (EXC 2147, project-id 390858490).
\end{acknowledgments}

\appendix
\section{Degeneracy from nearest neighbor hopping}
\label{app:degeneracy}
Consider a bipartite lattice with $2n$ atoms per unit cell and NN hopping only. Then, the lattice Hamiltonian can be decomposed into two spin-blocks of the form
\begin{equation}
    \mathcal H = \sum_{\vec{k},\sigma} \vec{c}_{\vec{k}\sigma}^\dagger H_\sigma(\vec{k}) \vec{c}_{\vec{k}\sigma},
\end{equation}
where the basis is ordered by sublattice, first $A$ and then $B$ so that $\vec{c}_{\vec{k}\sigma} = (c_{\vec{k}\sigma,A,1}, \dots c_{\vec{k}\sigma,A,n}, c_{\vec{k}\sigma,B,1}, \dots c_{\vec{k}\sigma,B,n})^T$ and
\begin{equation}
    H_{\sigma}(\vec{k})=
    \begin{pmatrix}
        -J\sigma\,\mathbb{1}_n&t(\vec{k})\\
        t(\vec{k})^\dagger&J\sigma\,\mathbb{1}_n
    \end{pmatrix}.
\end{equation}
Here $t(\vec{k})$ denotes the hopping matrix between the two different atom types. Since NN hopping only acts between sublattices $A$ and $B$, the hopping terms are entirely confined to the off-diagonal blocks of the spin blocks.
Note that, because of this construction, squaring a single spin block yields
\begin{equation}
    H_\sigma^2(\vec{k})=
    \begin{pmatrix}
        (J\sigma)^2\,\mathbb{1}_n+t(\vec{k})t(\vec{k})^\dagger&0\\
        0&(J\sigma)^2\,\mathbb{1}_n+t(\vec{k})^\dagger t(\vec{k})
    \end{pmatrix}.
\end{equation}
One can consequently prove spin degeneracy across the entire BZ by taking the square root of the eigenvalues of $H_\sigma^2(\vec{k})$,
\begin{equation}
    E_{\sigma,m,\pm}(\vec{k})=\pm\sqrt{J^2\sigma^2+\lambda_m(\vec{k})},
\end{equation}
where $\lambda_m(\vec{k})$ are the eigenvalues of $t(\vec{k})t(\vec{k})^\dagger$.
Since $\sigma^2=1$, we observe no spin dependence of the eigenvalues if only NN hopping is taken into account.

\section{Abelian Hyperbolic antiferromagnets}
\label{app:anti}
In this article, we denote the hopping terms with a phase as $E_{ijk\dots}=e^{ik_i+ik_j+ik_k+\dots}$ and use $E_{\bar{i}}=e^{-ik_i}$ \cite{Chen2023Aug}. In the cases where hoppings simplify into cosines, we use $C_{ijk\dots}=\cos\left( k_i+k_j+k_k+\dots \right)$ and $C_{\bar{i}}=\cos\left(-k_i\right)$. Note, that we only examine NNN hopping for simplicity.

\subsection{\texorpdfstring{$\{10,5\}$}{\{10,5\}} lattice}
\label{app:105}
For the $\{10,5\}$ lattice shown in Fig.~\ref{fig:antiferro}(b), we derive
\begin{equation}
    H_{\{10,5\},\sigma}(\vec{k})=
    \begin{pmatrix}
        -J\sigma+t_A(\vec{k})&0\\
        0&J\sigma+t_B(\vec{k})
    \end{pmatrix}
\end{equation}
with
\begin{equation}
    t_A(\vec{k})=t_B(\vec{k})=2t_2\left( C_{2\bar{1}} + C_{1\bar{2}3} + C_{2\bar{3}4} + C_{23} + C_{3\bar{4}} \right).
\end{equation}
The hopping terms are exactly the same for the two sublattices. It is thus trivial to show that $H_{\{10,5\}}$ is invariant under $\mathcal{P}\mathcal{T}$. Using the matrix form of the P operation
\begin{equation}
    M_P^{\{10,5\}}=\begin{pmatrix}
        1&-1&1&-1\\
        1&0&0&0\\
        0&1&0&0\\
        0&0&1&0
    \end{pmatrix},
\end{equation}
we find $P^5=-\mathbb{1}$ to show that $\mathcal{P}\mathcal{T}=P^5\mathcal{T}$ as expected.

\subsection{\texorpdfstring{$\{10,3\}$}{\{10,3\}} lattice}
\label{app:103}
For the $\{10,3\}$ lattice shown in Fig.~\ref{fig:antiferro}(c), we obtain
\begin{equation}
    H_{\{10,3\},\sigma}(\vec{k})=
    \begin{pmatrix}
        -J\sigma\,\mathbb{1}_5+t_A(\vec{k})&0\\
        0&J\sigma\,\mathbb{1}_5+t_B(\vec{k})
    \end{pmatrix}
\end{equation}
with
\begin{widetext}
\begin{equation}
    t_A(\vec{k})=t_B(\vec{k})^*=t_2
    \begin{pmatrix}
        0&1&E_1+E_{\bar{5}}&E_{1}+E_{2}&1\\
        1&0&1&E_{3}+E_{2}&E_{3}+E_{4}\\
        E_{\bar{1}}+E_{5}&1&0&1&E_{5}+E_{4}\\
        E_{\bar{1}}+E_{\bar{2}}&E_{\bar{3}}+E_{\bar{2}}&1&0&1\\
        1&E_{\bar{3}}+E_{\bar{4}}&E_{\bar{5}}+E_{\bar{4}}&1&0
    \end{pmatrix}.
\end{equation}
\end{widetext}
We once again trivially find $H_{\{10,3\}}(\vec{k},\sigma)=H_{\{10,3\}}(\vec{k},-\sigma)$ using $\mathcal{P}\mathcal{T}=P^5\mathcal{T}$, where $P$ is the same as for the $\{10,5\}$ lattice.

\subsection{\texorpdfstring{$\{8,3\}$}{\{8,3\}} lattice}
\label{app:83}
For the exceptional $\{8,3\}$ lattice shown in Fig.~\ref{fig:antiferro}(a), the tight-binding Hamiltonian takes the form
\begin{equation}
    H_{\{8,3\},\sigma}(\vec{k})=
    \begin{pmatrix}
        -J\sigma\,\mathbb{1}_8+t_A(\vec{k})&0\\
        0&J\sigma\,\mathbb{1}_8+t_B(\vec{k})
    \end{pmatrix}
\end{equation}
with 
\begin{equation}
    t_A(\vec{k})=t_2
    \begin{pmatrix}
    0&1&0&1&1&1&E_{\bar{4}}&E_{1}\\
    1&0&1&0&E_{3}&1&1&E_{2}\\
    0&1&0&1&E_{4}&E_{\bar{1}}&1&1\\
    1&0&1&0&1&E_{\bar{2}}&E_{\bar{3}}&1\\
    1&E_{\bar{3}}&E_{\bar{4}}&1&0&E_{\bar{4}\bar{1}}&0&E_{\bar{3}2}\\
    1&1&E_{1}&E_{2}&E_{41}&0&E_{\bar{3}2}&0\\
    E_{4}&1&1&E_{3}&0&E_{3\bar{2}}&0&E_{41}\\
    E_{\bar{1}}&E_{\bar{2}}&1&1&E_{3\bar{2}}&0&E_{\bar{4}\bar{1}}&0
    \end{pmatrix}
\end{equation}
and
\begin{equation}
    t_B(\vec{k})=t_2
    \begin{pmatrix}
    0&1&0&1&1&1&E_{1}&E_{2}\\
    1&0&1&0&E_{4}&1&1&E_{3}\\
    0&1&0&1&E_{\bar{1}}&E_{\bar{2}}&1&1\\
    1&0&1&0&1&E_{\bar{3}}&E_{\bar{4}}&1\\
    1&E_{\bar{4}}&E_{1}&1&0&E_{1\bar{2}}&0&E_{\bar{4}3}\\
    1&1&E_{2}&E_{3}&E_{\bar{1}2}&0&E_{3\bar{4}}&0\\
    E_{\bar{1}}&1&1&E_{4}&0&E_{\bar{3}4}&0&E_{\bar{1}2}\\
    E_{\bar{2}}&E_{\bar{3}}&1&1&E_{4\bar{3}}&0&E_{1\bar{2}}&0
    \end{pmatrix}.
\end{equation}

\section{Hyperbolic altermagnets}

In this appendix, we list the full Hamiltonians including NN hopping, together with the generators of the symmetries of the point groups of the $g=2$ lattices in the main text, which we find to be altermagnetic.

\subsection{\texorpdfstring{$\{8,4\}$}{\{8,4\}} lattice}
\label{app:84}

\begin{figure*}
    \centering
    \includegraphics[width=0.95\linewidth]{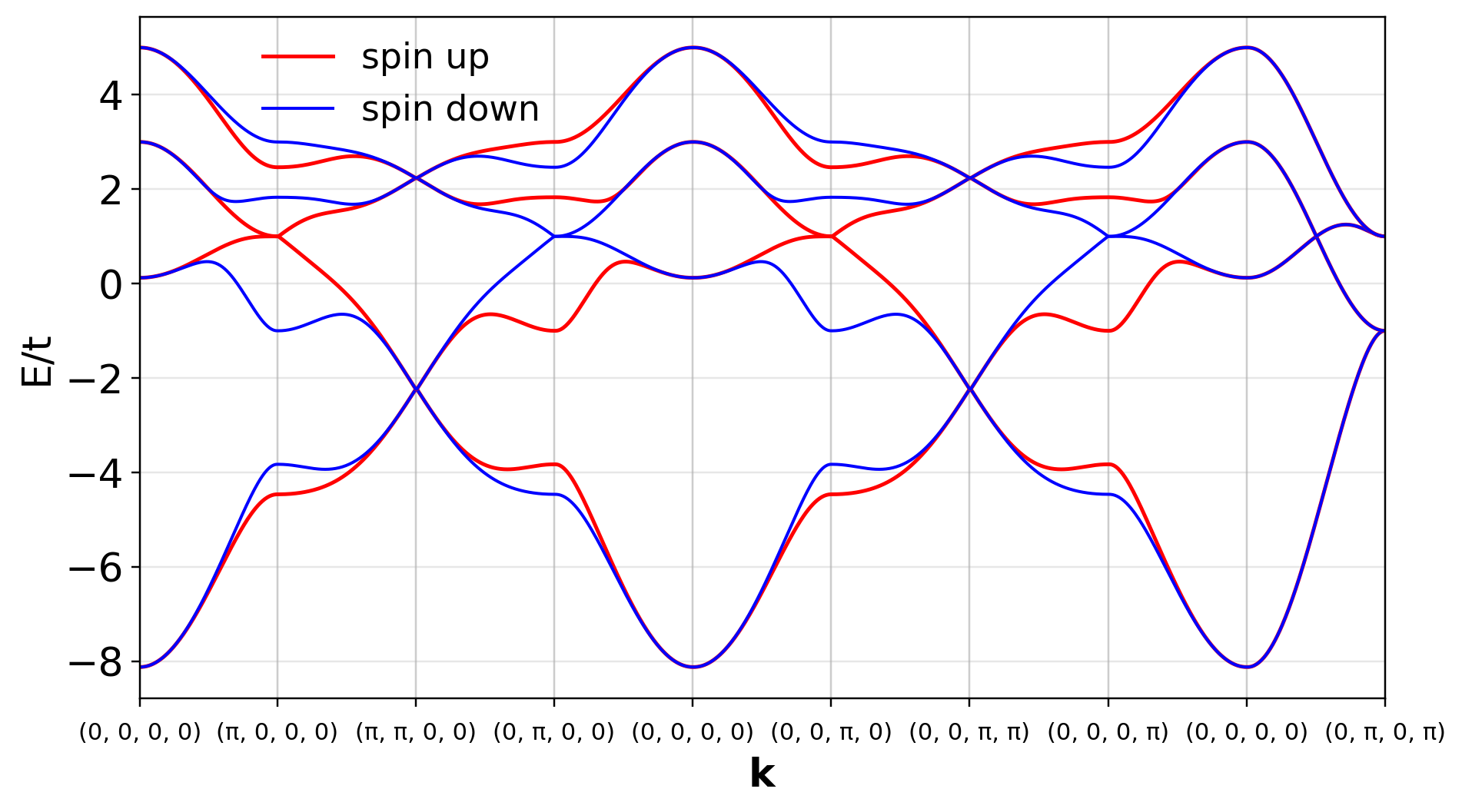}
    \caption{Example of a hyperbolic $4$D band structure using the $\{8,4\}$ lattice and Abelian hyperbolic band theory. Red/blue bands denote spin-up/spin-down bands respectively. In this plot, we enable both NN and NNN hopping with parameters set to $t=J=2t_2$.}
    \label{fig:bands}
\end{figure*}

The full Hamiltonian of the $\{8,4\}$ lattice shown in Fig.~\ref{fig:84} as the $g=2$ example for the $\{4g,4\}$ family of lattices is given by
\begin{equation}
    H_{\{8,4\},\sigma}(\vec{k})=
    \begin{pmatrix}
        -J\sigma\,\mathbb{1}_2+t_{NNN,A}(\vec{k})&t_{NN}(\vec{k})\\
        t_{NN}(\vec{k})^\dagger&J\sigma\,\mathbb{1}_2+t_{NNN,B}(\vec{k})
    \end{pmatrix}
\end{equation}
with
\begin{equation}
t_{NN}(\vec{k})=
-t\begin{pmatrix}
1+E_{1\bar{2}} & E_1+E_{\bar{4}}\\
1+E_{3\bar{2}} & 1+E_{3\bar{4}}
\end{pmatrix}
\end{equation}
and
\begin{equation}
t_{NNN,A/B}(\vec{k})=-t_2
\begin{pmatrix}
0 & t_{A/B}(\vec{k})\\
t_{A/B}(\vec{k})^\dagger & 0
\end{pmatrix},
\end{equation}
where
\begin{align}
t_A(\vec{k})&=1+E_{\bar{3}}+E_1+E_{1\bar{3}}+E_{2\bar{3}}+E_{14\bar{3}}+E_{1\bar{2}}+E_{\bar{4}},\\
t_B(\vec{k})&=1+E_{\bar{4}}+E_2+E_{2\bar{4}}+E_{3\bar{4}}+E_{2\bar{1}\bar{4}}+E_{2\bar{3}}+E_1.
\end{align}
An example of the resulting band structure is shown in Fig.~\ref{fig:bands}.
Below, we list the matrices of the generators of the symmetries of the $\{8,4\}$ lattice shown in Fig.~\ref{fig:84}. They are
\begin{widetext}
\begin{equation}
\begin{aligned}
M_a^{\{8,4\}} &=
\begin{pmatrix}
0 & 0 & 0 & -1\\
0 & 0 & -1 & 0\\
0 & -1 & 0 & 0\\
-1 & 0 & 0 & 0
\end{pmatrix},
&
M_b^{\{8,4\}} &=
\begin{pmatrix}
0 & 1 & -1 & 1\\
1 & 0 & -1 & 1\\
1 & -1 & 0 & 1\\
1 & -1 & 1 & 0
\end{pmatrix},
&
M_c^{\{8,4\}} &=
\begin{pmatrix}
1 & 0 & 0 & 0\\
0 & 0 & 0 & -1\\
0 & 0 & -1 & 0\\
0 & -1 & 0 & 0
\end{pmatrix},
\\[1em]
M_P^{\{8,4\}} &=
\begin{pmatrix}
0 & 0 & 0 & -1\\
1 & 0 & 0 & 0\\
0 & 1 & 0 & 0\\
0 & 0 & 1 & 0
\end{pmatrix},
&
M_Q^{\{8,4\}} &=
\begin{pmatrix}
0 & -1 & 1 & -1\\
1 & -1 & 1 & 0\\
1 & -1 & 0 & 1\\
1 & 0 & -1 & 1
\end{pmatrix},
&
M_R^{\{8,4\}} &=
\begin{pmatrix}
-1 & 1 & -1 & 0\\
-1 & 1 & 0 & -1\\
-1 & 0 & 1 & -1\\
0 & -1 & 1 & -1
\end{pmatrix}.
\end{aligned}
\end{equation}
\end{widetext}

\subsection{\texorpdfstring{$\{4,8\}$}{\{4,8\}} lattice}
\label{app:48}

In this section, we present the Hamiltonian of the exceptional $\{4,8\}$ lattice including NN hopping.
\begin{equation}
    H_{\{4,8\},\sigma}(\vec{k})=\mu\sigma_0 - J\sigma\sigma_z + t_{NN}(\vec{k}) + t_{NNN}(\vec{k})
\end{equation}
with
\begin{equation}
    t_{NN}(\vec{k})=-t
    \begin{pmatrix}
        0& t_{A\rightarrow B}(\vec{k})\\
        t_{A\rightarrow B}(\vec{k})^\dagger& 0\\
    \end{pmatrix}
\end{equation}
and
\begin{equation}
    t_{NNN}(\vec{k})=-t_2
    \begin{pmatrix}
        t_{A\rightarrow A}(\vec{k})& 0\\
        0& t_{B\rightarrow B}(\vec{k})\\
    \end{pmatrix},
\end{equation}
where
\begin{align}
    t_{A\rightarrow B}(\vec{k})
    &= 1+E_{\bar{1}}+E_{4}+E_{2\bar{3}}+E_{1\bar{3}}+E_{3}+E_{14\bar{3}}+E_{1\bar{2}},\\
    t_{A\rightarrow A}(\vec{k})
    &= 2\left( C_{1}+C_{2\bar{3}4}+C_{2\bar{1}}+C_{14} \right),\\
    t_{B\rightarrow B}(\vec{k})
    &= 2\left( C_{3}+C_{41\bar{2}}+C_{4\bar{3}}+C_{3\bar{2}} \right).
\end{align}

The matrices of the generators of the symmetries of the $\{4,8\}$ lattice are
\begin{widetext}
\begin{equation}
\begin{aligned}
M_a^{\{4,8\}} &=
\begin{pmatrix}
0 & 0 & -1 & 0\\
0 & -1 & 0 & 0\\
-1 & 0 & 0 & 0\\
0 & 0 & 0 & 1
\end{pmatrix},
&
M_b^{\{4,8\}} &=
\begin{pmatrix}
1 & 0 & 0 & 1\\
1 & -1 & 1 & 0\\
0 & 0 & 1 & -1\\
0 & 0 & 0 & -1
\end{pmatrix},
&
M_c^{\{4,8\}} &=
\begin{pmatrix}
1 & 0 & 0 & 0\\
0 & 0 & 0 & -1\\
0 & 0 & -1 & 0\\
0 & -1 & 0 & 0
\end{pmatrix},
\\[1em]
M_P^{\{4,8\}} &=
\begin{pmatrix}
0 & 0 & -1 & 0\\
0 & 0 & 0 & -1\\
1 & 0 & 0 & 0\\
0 & 1 & 0 & 0
\end{pmatrix},
&
M_Q^{\{4,8\}} &=
\begin{pmatrix}
1 & -1 & 0 & 0\\
1 & 0 & -1 & 1\\
0 & 1 & -1 & 0\\
0 & 1 & 0 & 0
\end{pmatrix},
&
M_R^{\{4,8\}} &=
\begin{pmatrix}
0 & 0 & -1 & 1\\
-1 & 1 & -1 & 0\\
-1 & 0 & 0 & -1\\
0 & 0 & 0 & -1
\end{pmatrix}.
\end{aligned}
\end{equation}
\end{widetext}


\begin{thebibliography}{56}%
\makeatletter
\providecommand \@ifxundefined [1]{%
 \@ifx{#1\undefined}
}%
\providecommand \@ifnum [1]{%
 \ifnum #1\expandafter \@firstoftwo
 \else \expandafter \@secondoftwo
 \fi
}%
\providecommand \@ifx [1]{%
 \ifx #1\expandafter \@firstoftwo
 \else \expandafter \@secondoftwo
 \fi
}%
\providecommand \natexlab [1]{#1}%
\providecommand \enquote  [1]{``#1''}%
\providecommand \bibnamefont  [1]{#1}%
\providecommand \bibfnamefont [1]{#1}%
\providecommand \citenamefont [1]{#1}%
\providecommand \href@noop [0]{\@secondoftwo}%
\providecommand \href [0]{\begingroup \@sanitize@url \@href}%
\providecommand \@href[1]{\@@startlink{#1}\@@href}%
\providecommand \@@href[1]{\endgroup#1\@@endlink}%
\providecommand \@sanitize@url [0]{\catcode `\\12\catcode `\$12\catcode
  `\&12\catcode `\#12\catcode `\^12\catcode `\_12\catcode `\%12\relax}%
\providecommand \@@startlink[1]{}%
\providecommand \@@endlink[0]{}%
\providecommand \url  [0]{\begingroup\@sanitize@url \@url }%
\providecommand \@url [1]{\endgroup\@href {#1}{\urlprefix }}%
\providecommand \urlprefix  [0]{URL }%
\providecommand \Eprint [0]{\href }%
\providecommand \doibase [0]{https://doi.org/}%
\providecommand \selectlanguage [0]{\@gobble}%
\providecommand \bibinfo  [0]{\@secondoftwo}%
\providecommand \bibfield  [0]{\@secondoftwo}%
\providecommand \translation [1]{[#1]}%
\providecommand \BibitemOpen [0]{}%
\providecommand \bibitemStop [0]{}%
\providecommand \bibitemNoStop [0]{.\EOS\space}%
\providecommand \EOS [0]{\spacefactor3000\relax}%
\providecommand \BibitemShut  [1]{\csname bibitem#1\endcsname}%
\let\auto@bib@innerbib\@empty
\bibitem [{\citenamefont {{\ifmmode\check{S}\else\v{S}\fi}mejkal}\ \emph
  {et~al.}(2022{\natexlab{a}})\citenamefont
  {{\ifmmode\check{S}\else\v{S}\fi}mejkal}, \citenamefont {Sinova},\ and\
  \citenamefont {Jungwirth}}]{Smejkal2022Sep}%
  \BibitemOpen
  \bibfield  {author} {\bibinfo {author} {\bibfnamefont {L.}~\bibnamefont
  {{\ifmmode\check{S}\else\v{S}\fi}mejkal}}, \bibinfo {author} {\bibfnamefont
  {J.}~\bibnamefont {Sinova}},\ and\ \bibinfo {author} {\bibfnamefont
  {T.}~\bibnamefont {Jungwirth}},\ }\bibfield  {title} {\bibinfo {title}
  {{Beyond Conventional Ferromagnetism and Antiferromagnetism: A Phase with
  Nonrelativistic Spin and Crystal Rotation Symmetry}},\ }\href
  {https://doi.org/10.1103/PhysRevX.12.031042} {\bibfield  {journal} {\bibinfo
  {journal} {Phys. Rev. X}\ }\textbf {\bibinfo {volume} {12}},\ \bibinfo
  {pages} {031042} (\bibinfo {year} {2022}{\natexlab{a}})}\BibitemShut
  {NoStop}%
\bibitem [{\citenamefont {Hayami}\ \emph {et~al.}(2019)\citenamefont {Hayami},
  \citenamefont {Yanagi},\ and\ \citenamefont {Kusunose}}]{Hayami2019Nov}%
  \BibitemOpen
  \bibfield  {author} {\bibinfo {author} {\bibfnamefont {S.}~\bibnamefont
  {Hayami}}, \bibinfo {author} {\bibfnamefont {Y.}~\bibnamefont {Yanagi}},\
  and\ \bibinfo {author} {\bibfnamefont {H.}~\bibnamefont {Kusunose}},\
  }\bibfield  {title} {\bibinfo {title} {{Momentum-Dependent Spin Splitting by
  Collinear Antiferromagnetic Ordering}},\ }\href
  {https://doi.org/10.7566/JPSJ.88.123702} {\bibfield  {journal} {\bibinfo
  {journal} {J. Phys. Soc. Jpn.}\ }\textbf {\bibinfo {volume} {88}},\ \bibinfo
  {pages} {123702} (\bibinfo {year} {2019})}\BibitemShut {NoStop}%
\bibitem [{\citenamefont {{\ifmmode\check{S}\else\v{S}\fi}mejkal}\ \emph
  {et~al.}(2020)\citenamefont {{\ifmmode\check{S}\else\v{S}\fi}mejkal},
  \citenamefont
  {Gonz{\ifmmode\acute{a}\else\'{a}\fi}lez-Hern{\ifmmode\acute{a}\else\'{a}\fi}ndez},
  \citenamefont {Jungwirth},\ and\ \citenamefont {Sinova}}]{Smejkal2020Jun}%
  \BibitemOpen
  \bibfield  {author} {\bibinfo {author} {\bibfnamefont {L.}~\bibnamefont
  {{\ifmmode\check{S}\else\v{S}\fi}mejkal}}, \bibinfo {author} {\bibfnamefont
  {R.}~\bibnamefont
  {Gonz{\ifmmode\acute{a}\else\'{a}\fi}lez-Hern{\ifmmode\acute{a}\else\'{a}\fi}ndez}},
  \bibinfo {author} {\bibfnamefont {T.}~\bibnamefont {Jungwirth}},\ and\
  \bibinfo {author} {\bibfnamefont {J.}~\bibnamefont {Sinova}},\ }\bibfield
  {title} {\bibinfo {title} {{Crystal time-reversal symmetry breaking and
  spontaneous Hall effect in collinear antiferromagnets}},\ }\href
  {https://doi.org/10.1126/sciadv.aaz8809} {\bibfield  {journal} {\bibinfo
  {journal} {Sci. Adv.}\ }\textbf {\bibinfo {volume} {6}},\ \bibinfo {pages}
  {eaaz8809} (\bibinfo {year} {2020})}\BibitemShut {NoStop}%
\bibitem [{\citenamefont {Yuan}\ \emph {et~al.}(2020)\citenamefont {Yuan},
  \citenamefont {Wang}, \citenamefont {Luo}, \citenamefont {Rashba},\ and\
  \citenamefont {Zunger}}]{Yuan2020Jul}%
  \BibitemOpen
  \bibfield  {author} {\bibinfo {author} {\bibfnamefont {L.-D.}\ \bibnamefont
  {Yuan}}, \bibinfo {author} {\bibfnamefont {Z.}~\bibnamefont {Wang}}, \bibinfo
  {author} {\bibfnamefont {J.-W.}\ \bibnamefont {Luo}}, \bibinfo {author}
  {\bibfnamefont {E.~I.}\ \bibnamefont {Rashba}},\ and\ \bibinfo {author}
  {\bibfnamefont {A.}~\bibnamefont {Zunger}},\ }\bibfield  {title} {\bibinfo
  {title} {{Giant momentum-dependent spin splitting in centrosymmetric low-$Z$
  antiferromagnets}},\ }\href {https://doi.org/10.1103/PhysRevB.102.014422}
  {\bibfield  {journal} {\bibinfo  {journal} {Phys. Rev. B}\ }\textbf {\bibinfo
  {volume} {102}},\ \bibinfo {pages} {014422} (\bibinfo {year}
  {2020})}\BibitemShut {NoStop}%
\bibitem [{\citenamefont {McClarty}\ and\ \citenamefont
  {Rau}(2024)}]{McClarty2024Apr}%
  \BibitemOpen
  \bibfield  {author} {\bibinfo {author} {\bibfnamefont {P.~A.}\ \bibnamefont
  {McClarty}}\ and\ \bibinfo {author} {\bibfnamefont {J.~G.}\ \bibnamefont
  {Rau}},\ }\bibfield  {title} {\bibinfo {title} {{Landau Theory of
  Altermagnetism}},\ }\href {https://doi.org/10.1103/PhysRevLett.132.176702}
  {\bibfield  {journal} {\bibinfo  {journal} {Phys. Rev. Lett.}\ }\textbf
  {\bibinfo {volume} {132}},\ \bibinfo {pages} {176702} (\bibinfo {year}
  {2024})}\BibitemShut {NoStop}%
\bibitem [{\citenamefont {Bhowal}\ and\ \citenamefont
  {Spaldin}(2024)}]{Bhowal2024Feb}%
  \BibitemOpen
  \bibfield  {author} {\bibinfo {author} {\bibfnamefont {S.}~\bibnamefont
  {Bhowal}}\ and\ \bibinfo {author} {\bibfnamefont {N.~A.}\ \bibnamefont
  {Spaldin}},\ }\bibfield  {title} {\bibinfo {title} {{Ferroically Ordered
  Magnetic Octupoles in $d$-Wave Altermagnets}},\ }\href
  {https://doi.org/10.1103/PhysRevX.14.011019} {\bibfield  {journal} {\bibinfo
  {journal} {Phys. Rev. X}\ }\textbf {\bibinfo {volume} {14}},\ \bibinfo
  {pages} {011019} (\bibinfo {year} {2024})}\BibitemShut {NoStop}%
\bibitem [{\citenamefont {{\ifmmode\check{S}\else\v{S}\fi}mejkal}\ \emph
  {et~al.}(2022{\natexlab{b}})\citenamefont
  {{\ifmmode\check{S}\else\v{S}\fi}mejkal}, \citenamefont {Sinova},\ and\
  \citenamefont {Jungwirth}}]{Smejkal2022Dec}%
  \BibitemOpen
  \bibfield  {author} {\bibinfo {author} {\bibfnamefont {L.}~\bibnamefont
  {{\ifmmode\check{S}\else\v{S}\fi}mejkal}}, \bibinfo {author} {\bibfnamefont
  {J.}~\bibnamefont {Sinova}},\ and\ \bibinfo {author} {\bibfnamefont
  {T.}~\bibnamefont {Jungwirth}},\ }\bibfield  {title} {\bibinfo {title}
  {{Emerging Research Landscape of Altermagnetism}},\ }\href
  {https://doi.org/10.1103/PhysRevX.12.040501} {\bibfield  {journal} {\bibinfo
  {journal} {Phys. Rev. X}\ }\textbf {\bibinfo {volume} {12}},\ \bibinfo
  {pages} {040501} (\bibinfo {year} {2022}{\natexlab{b}})}\BibitemShut
  {NoStop}%
\bibitem [{\citenamefont {{\ifmmode\check{S}\else\v{S}\fi}mejkal}\ \emph
  {et~al.}(2022{\natexlab{c}})\citenamefont
  {{\ifmmode\check{S}\else\v{S}\fi}mejkal}, \citenamefont {Hellenes},
  \citenamefont
  {Gonz{\ifmmode\acute{a}\else\'{a}\fi}lez-Hern{\ifmmode\acute{a}\else\'{a}\fi}ndez},
  \citenamefont {Sinova},\ and\ \citenamefont {Jungwirth}}]{Smejkal2022Feb}%
  \BibitemOpen
  \bibfield  {author} {\bibinfo {author} {\bibfnamefont {L.}~\bibnamefont
  {{\ifmmode\check{S}\else\v{S}\fi}mejkal}}, \bibinfo {author} {\bibfnamefont
  {A.~B.}\ \bibnamefont {Hellenes}}, \bibinfo {author} {\bibfnamefont
  {R.}~\bibnamefont
  {Gonz{\ifmmode\acute{a}\else\'{a}\fi}lez-Hern{\ifmmode\acute{a}\else\'{a}\fi}ndez}},
  \bibinfo {author} {\bibfnamefont {J.}~\bibnamefont {Sinova}},\ and\ \bibinfo
  {author} {\bibfnamefont {T.}~\bibnamefont {Jungwirth}},\ }\bibfield  {title}
  {\bibinfo {title} {{Giant and Tunneling Magnetoresistance in Unconventional
  Collinear Antiferromagnets with Nonrelativistic Spin-Momentum Coupling}},\
  }\href {https://doi.org/10.1103/PhysRevX.12.011028} {\bibfield  {journal}
  {\bibinfo  {journal} {Phys. Rev. X}\ }\textbf {\bibinfo {volume} {12}},\
  \bibinfo {pages} {011028} (\bibinfo {year} {2022}{\natexlab{c}})}\BibitemShut
  {NoStop}%
\bibitem [{\citenamefont {Fedchenko}\ \emph {et~al.}(2024)\citenamefont
  {Fedchenko}, \citenamefont {Min{\ifmmode\acute{a}\else\'{a}\fi}r},
  \citenamefont {Akashdeep}, \citenamefont {D{'}Souza}, \citenamefont
  {Vasilyev}, \citenamefont {Tkach}, \citenamefont {Odenbreit}, \citenamefont
  {Nguyen}, \citenamefont {Kutnyakhov}, \citenamefont {Wind}, \citenamefont
  {Wenthaus}, \citenamefont {Scholz}, \citenamefont {Rossnagel}, \citenamefont
  {Hoesch}, \citenamefont {Aeschlimann}, \citenamefont
  {Stadtm{\ifmmode\ddot{u}\else\"{u}\fi}ller}, \citenamefont
  {Kl{\ifmmode\ddot{a}\else\"{a}\fi}ui}, \citenamefont
  {Sch{\ifmmode\ddot{o}\else\"{o}\fi}nhense}, \citenamefont {Jungwirth},
  \citenamefont {Hellenes}, \citenamefont {Jakob}, \citenamefont
  {{\ifmmode\check{S}\else\v{S}\fi}mejkal}, \citenamefont {Sinova},\ and\
  \citenamefont {Elmers}}]{Fedchenko2024Jan}%
  \BibitemOpen
  \bibfield  {author} {\bibinfo {author} {\bibfnamefont {O.}~\bibnamefont
  {Fedchenko}}, \bibinfo {author} {\bibfnamefont {J.}~\bibnamefont
  {Min{\ifmmode\acute{a}\else\'{a}\fi}r}}, \bibinfo {author} {\bibfnamefont
  {A.}~\bibnamefont {Akashdeep}}, \bibinfo {author} {\bibfnamefont {S.~W.}\
  \bibnamefont {D{'}Souza}}, \bibinfo {author} {\bibfnamefont {D.}~\bibnamefont
  {Vasilyev}}, \bibinfo {author} {\bibfnamefont {O.}~\bibnamefont {Tkach}},
  \bibinfo {author} {\bibfnamefont {L.}~\bibnamefont {Odenbreit}}, \bibinfo
  {author} {\bibfnamefont {Q.}~\bibnamefont {Nguyen}}, \bibinfo {author}
  {\bibfnamefont {D.}~\bibnamefont {Kutnyakhov}}, \bibinfo {author}
  {\bibfnamefont {N.}~\bibnamefont {Wind}}, \bibinfo {author} {\bibfnamefont
  {L.}~\bibnamefont {Wenthaus}}, \bibinfo {author} {\bibfnamefont
  {M.}~\bibnamefont {Scholz}}, \bibinfo {author} {\bibfnamefont
  {K.}~\bibnamefont {Rossnagel}}, \bibinfo {author} {\bibfnamefont
  {M.}~\bibnamefont {Hoesch}}, \bibinfo {author} {\bibfnamefont
  {M.}~\bibnamefont {Aeschlimann}}, \bibinfo {author} {\bibfnamefont
  {B.}~\bibnamefont {Stadtm{\ifmmode\ddot{u}\else\"{u}\fi}ller}}, \bibinfo
  {author} {\bibfnamefont {M.}~\bibnamefont
  {Kl{\ifmmode\ddot{a}\else\"{a}\fi}ui}}, \bibinfo {author} {\bibfnamefont
  {G.}~\bibnamefont {Sch{\ifmmode\ddot{o}\else\"{o}\fi}nhense}}, \bibinfo
  {author} {\bibfnamefont {T.}~\bibnamefont {Jungwirth}}, \bibinfo {author}
  {\bibfnamefont {A.~B.}\ \bibnamefont {Hellenes}}, \bibinfo {author}
  {\bibfnamefont {G.}~\bibnamefont {Jakob}}, \bibinfo {author} {\bibfnamefont
  {L.}~\bibnamefont {{\ifmmode\check{S}\else\v{S}\fi}mejkal}}, \bibinfo
  {author} {\bibfnamefont {J.}~\bibnamefont {Sinova}},\ and\ \bibinfo {author}
  {\bibfnamefont {H.-J.}\ \bibnamefont {Elmers}},\ }\bibfield  {title}
  {\bibinfo {title} {{Observation of time-reversal symmetry breaking in the
  band structure of altermagnetic RuO$_2$}},\ }\href
  {https://doi.org/10.1126/sciadv.adj4883} {\bibfield  {journal} {\bibinfo
  {journal} {Sci. Adv.}\ }\textbf {\bibinfo {volume} {10}},\ \bibinfo {pages}
  {eadj4883} (\bibinfo {year} {2024})}\BibitemShut {NoStop}%
\bibitem [{\citenamefont {Liu}\ \emph {et~al.}(2024)\citenamefont {Liu},
  \citenamefont {Zhan}, \citenamefont {Li}, \citenamefont {Liu}, \citenamefont
  {Cheng}, \citenamefont {Shi}, \citenamefont {Deng}, \citenamefont {Zhang},
  \citenamefont {Li}, \citenamefont {Ding}, \citenamefont {Jiang},
  \citenamefont {Ye}, \citenamefont {Liu}, \citenamefont {Jiang}, \citenamefont
  {Wang}, \citenamefont {Li}, \citenamefont {Xie}, \citenamefont {Wang},
  \citenamefont {Qiao}, \citenamefont {Wen}, \citenamefont {Sun},\ and\
  \citenamefont {Shen}}]{Liu2024Oct}%
  \BibitemOpen
  \bibfield  {author} {\bibinfo {author} {\bibfnamefont {J.}~\bibnamefont
  {Liu}}, \bibinfo {author} {\bibfnamefont {J.}~\bibnamefont {Zhan}}, \bibinfo
  {author} {\bibfnamefont {T.}~\bibnamefont {Li}}, \bibinfo {author}
  {\bibfnamefont {J.}~\bibnamefont {Liu}}, \bibinfo {author} {\bibfnamefont
  {S.}~\bibnamefont {Cheng}}, \bibinfo {author} {\bibfnamefont
  {Y.}~\bibnamefont {Shi}}, \bibinfo {author} {\bibfnamefont {L.}~\bibnamefont
  {Deng}}, \bibinfo {author} {\bibfnamefont {M.}~\bibnamefont {Zhang}},
  \bibinfo {author} {\bibfnamefont {C.}~\bibnamefont {Li}}, \bibinfo {author}
  {\bibfnamefont {J.}~\bibnamefont {Ding}}, \bibinfo {author} {\bibfnamefont
  {Q.}~\bibnamefont {Jiang}}, \bibinfo {author} {\bibfnamefont
  {M.}~\bibnamefont {Ye}}, \bibinfo {author} {\bibfnamefont {Z.}~\bibnamefont
  {Liu}}, \bibinfo {author} {\bibfnamefont {Z.}~\bibnamefont {Jiang}}, \bibinfo
  {author} {\bibfnamefont {S.}~\bibnamefont {Wang}}, \bibinfo {author}
  {\bibfnamefont {Q.}~\bibnamefont {Li}}, \bibinfo {author} {\bibfnamefont
  {Y.}~\bibnamefont {Xie}}, \bibinfo {author} {\bibfnamefont {Y.}~\bibnamefont
  {Wang}}, \bibinfo {author} {\bibfnamefont {S.}~\bibnamefont {Qiao}}, \bibinfo
  {author} {\bibfnamefont {J.}~\bibnamefont {Wen}}, \bibinfo {author}
  {\bibfnamefont {Y.}~\bibnamefont {Sun}},\ and\ \bibinfo {author}
  {\bibfnamefont {D.}~\bibnamefont {Shen}},\ }\bibfield  {title} {\bibinfo
  {title} {{Absence of Altermagnetic Spin Splitting Character in Rutile Oxide
  ${\mathrm{RuO}}_{2}$}},\ }\href
  {https://doi.org/10.1103/PhysRevLett.133.176401} {\bibfield  {journal}
  {\bibinfo  {journal} {Phys. Rev. Lett.}\ }\textbf {\bibinfo {volume} {133}},\
  \bibinfo {pages} {176401} (\bibinfo {year} {2024})}\BibitemShut {NoStop}%
\bibitem [{\citenamefont {Mazin}(2023)}]{Mazin2023Mar}%
  \BibitemOpen
  \bibfield  {author} {\bibinfo {author} {\bibfnamefont {I.~I.}\ \bibnamefont
  {Mazin}},\ }\bibfield  {title} {\bibinfo {title} {{Altermagnetism in MnTe:
  Origin, predicted manifestations, and routes to detwinning}},\ }\href
  {https://doi.org/10.1103/PhysRevB.107.L100418} {\bibfield  {journal}
  {\bibinfo  {journal} {Phys. Rev. B}\ }\textbf {\bibinfo {volume} {107}},\
  \bibinfo {pages} {L100418} (\bibinfo {year} {2023})}\BibitemShut {NoStop}%
\bibitem [{\citenamefont {Lee}\ \emph {et~al.}(2024)\citenamefont {Lee},
  \citenamefont {Lee}, \citenamefont {Jung}, \citenamefont {Jung},
  \citenamefont {Kim}, \citenamefont {Lee}, \citenamefont {Seok}, \citenamefont
  {Kim}, \citenamefont {Park}, \citenamefont
  {{\ifmmode\check{S}\else\v{S}\fi}mejkal}, \citenamefont {Kang},\ and\
  \citenamefont {Kim}}]{Lee2024Jan}%
  \BibitemOpen
  \bibfield  {author} {\bibinfo {author} {\bibfnamefont {S.}~\bibnamefont
  {Lee}}, \bibinfo {author} {\bibfnamefont {S.}~\bibnamefont {Lee}}, \bibinfo
  {author} {\bibfnamefont {S.}~\bibnamefont {Jung}}, \bibinfo {author}
  {\bibfnamefont {J.}~\bibnamefont {Jung}}, \bibinfo {author} {\bibfnamefont
  {D.}~\bibnamefont {Kim}}, \bibinfo {author} {\bibfnamefont {Y.}~\bibnamefont
  {Lee}}, \bibinfo {author} {\bibfnamefont {B.}~\bibnamefont {Seok}}, \bibinfo
  {author} {\bibfnamefont {J.}~\bibnamefont {Kim}}, \bibinfo {author}
  {\bibfnamefont {B.~G.}\ \bibnamefont {Park}}, \bibinfo {author}
  {\bibfnamefont {L.}~\bibnamefont {{\ifmmode\check{S}\else\v{S}\fi}mejkal}},
  \bibinfo {author} {\bibfnamefont {C.-J.}\ \bibnamefont {Kang}},\ and\
  \bibinfo {author} {\bibfnamefont {C.}~\bibnamefont {Kim}},\ }\bibfield
  {title} {\bibinfo {title} {{Broken Kramers Degeneracy in Altermagnetic
  MnTe}},\ }\href {https://doi.org/10.1103/PhysRevLett.132.036702} {\bibfield
  {journal} {\bibinfo  {journal} {Phys. Rev. Lett.}\ }\textbf {\bibinfo
  {volume} {132}},\ \bibinfo {pages} {036702} (\bibinfo {year}
  {2024})}\BibitemShut {NoStop}%
\bibitem [{\citenamefont {Reimers}\ \emph {et~al.}(2024)\citenamefont
  {Reimers}, \citenamefont {Odenbreit}, \citenamefont
  {{\ifmmode\check{S}\else\v{S}\fi}mejkal}, \citenamefont {Strocov},
  \citenamefont {Constantinou}, \citenamefont {Hellenes}, \citenamefont
  {Jaeschke~Ubiergo}, \citenamefont {Campos}, \citenamefont {Bharadwaj},
  \citenamefont {Chakraborty}, \citenamefont {Denneulin}, \citenamefont {Shi},
  \citenamefont {Dunin-Borkowski}, \citenamefont {Das}, \citenamefont
  {Kl{\ifmmode\ddot{a}\else\"{a}\fi}ui}, \citenamefont {Sinova},\ and\
  \citenamefont {Jourdan}}]{Reimers2024Mar}%
  \BibitemOpen
  \bibfield  {author} {\bibinfo {author} {\bibfnamefont {S.}~\bibnamefont
  {Reimers}}, \bibinfo {author} {\bibfnamefont {L.}~\bibnamefont {Odenbreit}},
  \bibinfo {author} {\bibfnamefont {L.}~\bibnamefont
  {{\ifmmode\check{S}\else\v{S}\fi}mejkal}}, \bibinfo {author} {\bibfnamefont
  {V.~N.}\ \bibnamefont {Strocov}}, \bibinfo {author} {\bibfnamefont
  {P.}~\bibnamefont {Constantinou}}, \bibinfo {author} {\bibfnamefont {A.~B.}\
  \bibnamefont {Hellenes}}, \bibinfo {author} {\bibfnamefont {R.}~\bibnamefont
  {Jaeschke~Ubiergo}}, \bibinfo {author} {\bibfnamefont {W.~H.}\ \bibnamefont
  {Campos}}, \bibinfo {author} {\bibfnamefont {V.~K.}\ \bibnamefont
  {Bharadwaj}}, \bibinfo {author} {\bibfnamefont {A.}~\bibnamefont
  {Chakraborty}}, \bibinfo {author} {\bibfnamefont {T.}~\bibnamefont
  {Denneulin}}, \bibinfo {author} {\bibfnamefont {W.}~\bibnamefont {Shi}},
  \bibinfo {author} {\bibfnamefont {R.~E.}\ \bibnamefont {Dunin-Borkowski}},
  \bibinfo {author} {\bibfnamefont {S.}~\bibnamefont {Das}}, \bibinfo {author}
  {\bibfnamefont {M.}~\bibnamefont {Kl{\ifmmode\ddot{a}\else\"{a}\fi}ui}},
  \bibinfo {author} {\bibfnamefont {J.}~\bibnamefont {Sinova}},\ and\ \bibinfo
  {author} {\bibfnamefont {M.}~\bibnamefont {Jourdan}},\ }\bibfield  {title}
  {\bibinfo {title} {{Direct observation of altermagnetic band splitting in
  CrSb thin films}},\ }\href {https://doi.org/10.1038/s41467-024-46476-5}
  {\bibfield  {journal} {\bibinfo  {journal} {Nat. Commun.}\ }\textbf {\bibinfo
  {volume} {15}},\ \bibinfo {pages} {2116} (\bibinfo {year}
  {2024})}\BibitemShut {NoStop}%
\bibitem [{\citenamefont {Ding}\ \emph {et~al.}(2024)\citenamefont {Ding},
  \citenamefont {Jiang}, \citenamefont {Chen}, \citenamefont {Tao},
  \citenamefont {Liu}, \citenamefont {Li}, \citenamefont {Liu}, \citenamefont
  {Sun}, \citenamefont {Cheng}, \citenamefont {Liu}, \citenamefont {Yang},
  \citenamefont {Zhang}, \citenamefont {Deng}, \citenamefont {Jing},
  \citenamefont {Huang}, \citenamefont {Shi}, \citenamefont {Ye}, \citenamefont
  {Qiao}, \citenamefont {Wang}, \citenamefont {Guo}, \citenamefont {Feng},\
  and\ \citenamefont {Shen}}]{Ding2024Nov}%
  \BibitemOpen
  \bibfield  {author} {\bibinfo {author} {\bibfnamefont {J.}~\bibnamefont
  {Ding}}, \bibinfo {author} {\bibfnamefont {Z.}~\bibnamefont {Jiang}},
  \bibinfo {author} {\bibfnamefont {X.}~\bibnamefont {Chen}}, \bibinfo {author}
  {\bibfnamefont {Z.}~\bibnamefont {Tao}}, \bibinfo {author} {\bibfnamefont
  {Z.}~\bibnamefont {Liu}}, \bibinfo {author} {\bibfnamefont {T.}~\bibnamefont
  {Li}}, \bibinfo {author} {\bibfnamefont {J.}~\bibnamefont {Liu}}, \bibinfo
  {author} {\bibfnamefont {J.}~\bibnamefont {Sun}}, \bibinfo {author}
  {\bibfnamefont {J.}~\bibnamefont {Cheng}}, \bibinfo {author} {\bibfnamefont
  {J.}~\bibnamefont {Liu}}, \bibinfo {author} {\bibfnamefont {Y.}~\bibnamefont
  {Yang}}, \bibinfo {author} {\bibfnamefont {R.}~\bibnamefont {Zhang}},
  \bibinfo {author} {\bibfnamefont {L.}~\bibnamefont {Deng}}, \bibinfo {author}
  {\bibfnamefont {W.}~\bibnamefont {Jing}}, \bibinfo {author} {\bibfnamefont
  {Y.}~\bibnamefont {Huang}}, \bibinfo {author} {\bibfnamefont
  {Y.}~\bibnamefont {Shi}}, \bibinfo {author} {\bibfnamefont {M.}~\bibnamefont
  {Ye}}, \bibinfo {author} {\bibfnamefont {S.}~\bibnamefont {Qiao}}, \bibinfo
  {author} {\bibfnamefont {Y.}~\bibnamefont {Wang}}, \bibinfo {author}
  {\bibfnamefont {Y.}~\bibnamefont {Guo}}, \bibinfo {author} {\bibfnamefont
  {D.}~\bibnamefont {Feng}},\ and\ \bibinfo {author} {\bibfnamefont
  {D.}~\bibnamefont {Shen}},\ }\bibfield  {title} {\bibinfo {title} {{Large
  Band Splitting in $g$-Wave Altermagnet CrSb}},\ }\href
  {https://doi.org/10.1103/PhysRevLett.133.206401} {\bibfield  {journal}
  {\bibinfo  {journal} {Phys. Rev. Lett.}\ }\textbf {\bibinfo {volume} {133}},\
  \bibinfo {pages} {206401} (\bibinfo {year} {2024})}\BibitemShut {NoStop}%
\bibitem [{\citenamefont {Reichlova}\ \emph {et~al.}(2024)\citenamefont
  {Reichlova}, \citenamefont {Lopes~Seeger}, \citenamefont
  {Gonz{\ifmmode\acute{a}\else\'{a}\fi}lez-Hern{\ifmmode\acute{a}\else\'{a}\fi}ndez},
  \citenamefont {Kounta}, \citenamefont {Schlitz}, \citenamefont {Kriegner},
  \citenamefont {Ritzinger}, \citenamefont {Lammel}, \citenamefont
  {Leivisk{\ifmmode\ddot{a}\else\"{a}\fi}}, \citenamefont {Birk~Hellenes},
  \citenamefont {Olejn{\ifmmode\acute{\imath}\else\'{\i}\fi}k}, \citenamefont
  {Pet{\ifmmode\check{r}\else\v{r}\fi}i{\ifmmode\check{c}\else\v{c}\fi}ek},
  \citenamefont {Dole{\ifmmode\check{z}\else\v{z}\fi}al}, \citenamefont
  {Horak}, \citenamefont {Schmoranzerova}, \citenamefont {Badura},
  \citenamefont {Bertaina}, \citenamefont {Thomas}, \citenamefont {Baltz},
  \citenamefont {Michez}, \citenamefont {Sinova}, \citenamefont {Goennenwein},
  \citenamefont {Jungwirth},\ and\ \citenamefont
  {{\ifmmode\check{S}\else\v{S}\fi}mejkal}}]{Reichlova2024Jun}%
  \BibitemOpen
  \bibfield  {author} {\bibinfo {author} {\bibfnamefont {H.}~\bibnamefont
  {Reichlova}}, \bibinfo {author} {\bibfnamefont {R.}~\bibnamefont
  {Lopes~Seeger}}, \bibinfo {author} {\bibfnamefont {R.}~\bibnamefont
  {Gonz{\ifmmode\acute{a}\else\'{a}\fi}lez-Hern{\ifmmode\acute{a}\else\'{a}\fi}ndez}},
  \bibinfo {author} {\bibfnamefont {I.}~\bibnamefont {Kounta}}, \bibinfo
  {author} {\bibfnamefont {R.}~\bibnamefont {Schlitz}}, \bibinfo {author}
  {\bibfnamefont {D.}~\bibnamefont {Kriegner}}, \bibinfo {author}
  {\bibfnamefont {P.}~\bibnamefont {Ritzinger}}, \bibinfo {author}
  {\bibfnamefont {M.}~\bibnamefont {Lammel}}, \bibinfo {author} {\bibfnamefont
  {M.}~\bibnamefont {Leivisk{\ifmmode\ddot{a}\else\"{a}\fi}}}, \bibinfo
  {author} {\bibfnamefont {A.}~\bibnamefont {Birk~Hellenes}}, \bibinfo {author}
  {\bibfnamefont {K.}~\bibnamefont
  {Olejn{\ifmmode\acute{\imath}\else\'{\i}\fi}k}}, \bibinfo {author}
  {\bibfnamefont {V.}~\bibnamefont
  {Pet{\ifmmode\check{r}\else\v{r}\fi}i{\ifmmode\check{c}\else\v{c}\fi}ek}},
  \bibinfo {author} {\bibfnamefont {P.}~\bibnamefont
  {Dole{\ifmmode\check{z}\else\v{z}\fi}al}}, \bibinfo {author} {\bibfnamefont
  {L.}~\bibnamefont {Horak}}, \bibinfo {author} {\bibfnamefont
  {E.}~\bibnamefont {Schmoranzerova}}, \bibinfo {author} {\bibfnamefont
  {A.}~\bibnamefont {Badura}}, \bibinfo {author} {\bibfnamefont
  {S.}~\bibnamefont {Bertaina}}, \bibinfo {author} {\bibfnamefont
  {A.}~\bibnamefont {Thomas}}, \bibinfo {author} {\bibfnamefont
  {V.}~\bibnamefont {Baltz}}, \bibinfo {author} {\bibfnamefont
  {L.}~\bibnamefont {Michez}}, \bibinfo {author} {\bibfnamefont
  {J.}~\bibnamefont {Sinova}}, \bibinfo {author} {\bibfnamefont {S.~T.~B.}\
  \bibnamefont {Goennenwein}}, \bibinfo {author} {\bibfnamefont
  {T.}~\bibnamefont {Jungwirth}},\ and\ \bibinfo {author} {\bibfnamefont
  {L.}~\bibnamefont {{\ifmmode\check{S}\else\v{S}\fi}mejkal}},\ }\bibfield
  {title} {\bibinfo {title} {{Observation of a spontaneous anomalous Hall
  response in the Mn5Si3 d-wave altermagnet candidate}},\ }\href
  {https://doi.org/10.1038/s41467-024-48493-w} {\bibfield  {journal} {\bibinfo
  {journal} {Nat. Commun.}\ }\textbf {\bibinfo {volume} {15}},\ \bibinfo
  {pages} {4961} (\bibinfo {year} {2024})}\BibitemShut {NoStop}%
\bibitem [{\citenamefont {Rial}\ \emph {et~al.}(2024)\citenamefont {Rial},
  \citenamefont {Leivisk{\ifmmode\ddot{a}\else\"{a}\fi}}, \citenamefont
  {Skobjin}, \citenamefont {Bad'ura}, \citenamefont {Gaudin}, \citenamefont
  {Disdier}, \citenamefont {Schlitz}, \citenamefont {Kounta}, \citenamefont
  {Beckert}, \citenamefont {Kriegner}, \citenamefont {Thomas}, \citenamefont
  {Schmoranzerov{\ifmmode\acute{a}\else\'{a}\fi}}, \citenamefont
  {{\ifmmode\check{S}\else\v{S}\fi}mejkal}, \citenamefont {Sinova},
  \citenamefont {Jungwirth}, \citenamefont {Michez}, \citenamefont
  {Reichlov{\ifmmode\acute{a}\else\'{a}\fi}}, \citenamefont {Goennenwein},
  \citenamefont {Gomonay},\ and\ \citenamefont {Baltz}}]{Rial2024Dec}%
  \BibitemOpen
  \bibfield  {author} {\bibinfo {author} {\bibfnamefont {J.}~\bibnamefont
  {Rial}}, \bibinfo {author} {\bibfnamefont {M.}~\bibnamefont
  {Leivisk{\ifmmode\ddot{a}\else\"{a}\fi}}}, \bibinfo {author} {\bibfnamefont
  {G.}~\bibnamefont {Skobjin}}, \bibinfo {author} {\bibfnamefont
  {A.}~\bibnamefont {Bad'ura}}, \bibinfo {author} {\bibfnamefont
  {G.}~\bibnamefont {Gaudin}}, \bibinfo {author} {\bibfnamefont
  {F.}~\bibnamefont {Disdier}}, \bibinfo {author} {\bibfnamefont
  {R.}~\bibnamefont {Schlitz}}, \bibinfo {author} {\bibfnamefont
  {I.}~\bibnamefont {Kounta}}, \bibinfo {author} {\bibfnamefont
  {S.}~\bibnamefont {Beckert}}, \bibinfo {author} {\bibfnamefont
  {D.}~\bibnamefont {Kriegner}}, \bibinfo {author} {\bibfnamefont
  {A.}~\bibnamefont {Thomas}}, \bibinfo {author} {\bibfnamefont
  {E.}~\bibnamefont {Schmoranzerov{\ifmmode\acute{a}\else\'{a}\fi}}}, \bibinfo
  {author} {\bibfnamefont {L.}~\bibnamefont
  {{\ifmmode\check{S}\else\v{S}\fi}mejkal}}, \bibinfo {author} {\bibfnamefont
  {J.}~\bibnamefont {Sinova}}, \bibinfo {author} {\bibfnamefont
  {T.}~\bibnamefont {Jungwirth}}, \bibinfo {author} {\bibfnamefont
  {L.}~\bibnamefont {Michez}}, \bibinfo {author} {\bibfnamefont
  {H.}~\bibnamefont {Reichlov{\ifmmode\acute{a}\else\'{a}\fi}}}, \bibinfo
  {author} {\bibfnamefont {S.~T.~B.}\ \bibnamefont {Goennenwein}}, \bibinfo
  {author} {\bibfnamefont {O.}~\bibnamefont {Gomonay}},\ and\ \bibinfo {author}
  {\bibfnamefont {V.}~\bibnamefont {Baltz}},\ }\bibfield  {title} {\bibinfo
  {title} {{Altermagnetic variants in thin films of
  $\mathrm{M}{\mathrm{n}}_{5}\mathrm{S}{\mathrm{i}}_{3}$}},\ }\href
  {https://doi.org/10.1103/PhysRevB.110.L220411} {\bibfield  {journal}
  {\bibinfo  {journal} {Phys. Rev. B}\ }\textbf {\bibinfo {volume} {110}},\
  \bibinfo {pages} {L220411} (\bibinfo {year} {2024})}\BibitemShut {NoStop}%
\bibitem [{\citenamefont {Wei}\ \emph {et~al.}(2025)\citenamefont {Wei},
  \citenamefont {Li}, \citenamefont {Hatt}, \citenamefont {Huai}, \citenamefont
  {Liu}, \citenamefont {Singh}, \citenamefont {Kim}, \citenamefont {Fernandes},
  \citenamefont {Cardon}, \citenamefont {Zhao}, \citenamefont {Tran},
  \citenamefont {Frandsen}, \citenamefont {Burch}, \citenamefont {Liu},\ and\
  \citenamefont {Ji}}]{LAOMNSESource}%
  \BibitemOpen
  \bibfield  {author} {\bibinfo {author} {\bibfnamefont {C.-C.}\ \bibnamefont
  {Wei}}, \bibinfo {author} {\bibfnamefont {X.}~\bibnamefont {Li}}, \bibinfo
  {author} {\bibfnamefont {S.}~\bibnamefont {Hatt}}, \bibinfo {author}
  {\bibfnamefont {X.}~\bibnamefont {Huai}}, \bibinfo {author} {\bibfnamefont
  {J.}~\bibnamefont {Liu}}, \bibinfo {author} {\bibfnamefont {B.}~\bibnamefont
  {Singh}}, \bibinfo {author} {\bibfnamefont {K.-M.}\ \bibnamefont {Kim}},
  \bibinfo {author} {\bibfnamefont {R.~M.}\ \bibnamefont {Fernandes}}, \bibinfo
  {author} {\bibfnamefont {P.}~\bibnamefont {Cardon}}, \bibinfo {author}
  {\bibfnamefont {L.}~\bibnamefont {Zhao}}, \bibinfo {author} {\bibfnamefont
  {T.~T.}\ \bibnamefont {Tran}}, \bibinfo {author} {\bibfnamefont {B.~A.}\
  \bibnamefont {Frandsen}}, \bibinfo {author} {\bibfnamefont {K.~S.}\
  \bibnamefont {Burch}}, \bibinfo {author} {\bibfnamefont {F.}~\bibnamefont
  {Liu}},\ and\ \bibinfo {author} {\bibfnamefont {H.}~\bibnamefont {Ji}},\
  }\bibfield  {title} {\bibinfo {title}
  {{${\mathrm{La}}_{2}{\mathrm{O}}_{3}{\mathrm{Mn}}_{2}{\mathrm{Se}}_{2}$: A
  correlated insulating layered d-wave altermagnet}},\ }\href
  {https://doi.org/10.1103/PhysRevMaterials.9.024402} {\bibfield  {journal}
  {\bibinfo  {journal} {Phys. Rev. Mater.}\ }\textbf {\bibinfo {volume} {9}},\
  \bibinfo {pages} {024402} (\bibinfo {year} {2025})}\BibitemShut {NoStop}%
\bibitem [{\citenamefont {Garcia-Gassull}\ \emph {et~al.}(2026)\citenamefont
  {Garcia-Gassull}, \citenamefont {Razpopov}, \citenamefont {Stavropoulos},
  \citenamefont {Mazin},\ and\ \citenamefont {Valent{\'i}}}]{LAOMNSESource2}%
  \BibitemOpen
  \bibfield  {author} {\bibinfo {author} {\bibfnamefont {L.}~\bibnamefont
  {Garcia-Gassull}}, \bibinfo {author} {\bibfnamefont {A.}~\bibnamefont
  {Razpopov}}, \bibinfo {author} {\bibfnamefont {P.~P.}\ \bibnamefont
  {Stavropoulos}}, \bibinfo {author} {\bibfnamefont {I.~I.}\ \bibnamefont
  {Mazin}},\ and\ \bibinfo {author} {\bibfnamefont {R.}~\bibnamefont
  {Valent{\'i}}},\ }\bibfield  {title} {\bibinfo {title} {{Microscopic origin
  of the magnetic interactions and their experimental signatures in
  altermagnetic
  ${\mathrm{La}}_{2}{\mathrm{O}}_{3}{\mathrm{Mn}}_{2}{\mathrm{Se}}_{2}$}},\
  }\href {https://doi.org/10.1038/s44306-025-00125-9} {\bibfield  {journal}
  {\bibinfo  {journal} {npj Spintronics}\ }\textbf {\bibinfo {volume} {4}},\
  \bibinfo {pages} {9} (\bibinfo {year} {2026})}\BibitemShut {NoStop}%
\bibitem [{\citenamefont {Krempask{\ifmmode\acute{y}\else\'{y}\fi}}\ \emph
  {et~al.}(2024)\citenamefont {Krempask{\ifmmode\acute{y}\else\'{y}\fi}},
  \citenamefont {{\ifmmode\check{S}\else\v{S}\fi}mejkal}, \citenamefont
  {D{'}Souza}, \citenamefont {Hajlaoui}, \citenamefont {Springholz},
  \citenamefont
  {Uhl{\ifmmode\acute{\imath}\else\'{\i}\fi}{\ifmmode\check{r}\else\v{r}\fi}ov{\ifmmode\acute{a}\else\'{a}\fi}},
  \citenamefont {Alarab}, \citenamefont {Constantinou}, \citenamefont
  {Strocov}, \citenamefont {Usanov}, \citenamefont {Pudelko}, \citenamefont
  {Gonz{\ifmmode\acute{a}\else\'{a}\fi}lez-Hern{\ifmmode\acute{a}\else\'{a}\fi}ndez},
  \citenamefont {Birk~Hellenes}, \citenamefont {Jansa}, \citenamefont
  {Reichlov{\ifmmode\acute{a}\else\'{a}\fi}}, \citenamefont
  {{\ifmmode\check{S}\else\v{S}\fi}ob{\ifmmode\acute{a}\else\'{a}\fi}{\ifmmode\check{n}\else\v{n}\fi}},
  \citenamefont {Gonzalez~Betancourt}, \citenamefont {Wadley}, \citenamefont
  {Sinova}, \citenamefont {Kriegner}, \citenamefont
  {Min{\ifmmode\acute{a}\else\'{a}\fi}r}, \citenamefont {Dil},\ and\
  \citenamefont {Jungwirth}}]{Krempasky2024Feb}%
  \BibitemOpen
  \bibfield  {author} {\bibinfo {author} {\bibfnamefont {J.}~\bibnamefont
  {Krempask{\ifmmode\acute{y}\else\'{y}\fi}}}, \bibinfo {author} {\bibfnamefont
  {L.}~\bibnamefont {{\ifmmode\check{S}\else\v{S}\fi}mejkal}}, \bibinfo
  {author} {\bibfnamefont {S.~W.}\ \bibnamefont {D{'}Souza}}, \bibinfo {author}
  {\bibfnamefont {M.}~\bibnamefont {Hajlaoui}}, \bibinfo {author}
  {\bibfnamefont {G.}~\bibnamefont {Springholz}}, \bibinfo {author}
  {\bibfnamefont {K.}~\bibnamefont
  {Uhl{\ifmmode\acute{\imath}\else\'{\i}\fi}{\ifmmode\check{r}\else\v{r}\fi}ov{\ifmmode\acute{a}\else\'{a}\fi}}},
  \bibinfo {author} {\bibfnamefont {F.}~\bibnamefont {Alarab}}, \bibinfo
  {author} {\bibfnamefont {P.~C.}\ \bibnamefont {Constantinou}}, \bibinfo
  {author} {\bibfnamefont {V.}~\bibnamefont {Strocov}}, \bibinfo {author}
  {\bibfnamefont {D.}~\bibnamefont {Usanov}}, \bibinfo {author} {\bibfnamefont
  {W.~R.}\ \bibnamefont {Pudelko}}, \bibinfo {author} {\bibfnamefont
  {R.}~\bibnamefont
  {Gonz{\ifmmode\acute{a}\else\'{a}\fi}lez-Hern{\ifmmode\acute{a}\else\'{a}\fi}ndez}},
  \bibinfo {author} {\bibfnamefont {A.}~\bibnamefont {Birk~Hellenes}}, \bibinfo
  {author} {\bibfnamefont {Z.}~\bibnamefont {Jansa}}, \bibinfo {author}
  {\bibfnamefont {H.}~\bibnamefont {Reichlov{\ifmmode\acute{a}\else\'{a}\fi}}},
  \bibinfo {author} {\bibfnamefont {Z.}~\bibnamefont
  {{\ifmmode\check{S}\else\v{S}\fi}ob{\ifmmode\acute{a}\else\'{a}\fi}{\ifmmode\check{n}\else\v{n}\fi}}},
  \bibinfo {author} {\bibfnamefont {R.~D.}\ \bibnamefont
  {Gonzalez~Betancourt}}, \bibinfo {author} {\bibfnamefont {P.}~\bibnamefont
  {Wadley}}, \bibinfo {author} {\bibfnamefont {J.}~\bibnamefont {Sinova}},
  \bibinfo {author} {\bibfnamefont {D.}~\bibnamefont {Kriegner}}, \bibinfo
  {author} {\bibfnamefont {J.}~\bibnamefont
  {Min{\ifmmode\acute{a}\else\'{a}\fi}r}}, \bibinfo {author} {\bibfnamefont
  {J.~H.}\ \bibnamefont {Dil}},\ and\ \bibinfo {author} {\bibfnamefont
  {T.}~\bibnamefont {Jungwirth}},\ }\bibfield  {title} {\bibinfo {title}
  {{Altermagnetic lifting of Kramers spin degeneracy}},\ }\href
  {https://doi.org/10.1038/s41586-023-06907-7} {\bibfield  {journal} {\bibinfo
  {journal} {Nature}\ }\textbf {\bibinfo {volume} {626}},\ \bibinfo {pages}
  {517} (\bibinfo {year} {2024})}\BibitemShut {NoStop}%
\bibitem [{\citenamefont {Han}\ \emph {et~al.}(2024)\citenamefont {Han},
  \citenamefont {Fu}, \citenamefont {Peng}, \citenamefont {Cheng},
  \citenamefont {Dai}, \citenamefont {Liu}, \citenamefont {Li}, \citenamefont
  {Zhang}, \citenamefont {Zhu}, \citenamefont {Bai}, \citenamefont {Zhou},
  \citenamefont {Liang}, \citenamefont {Chen}, \citenamefont {Wang},
  \citenamefont {Chen}, \citenamefont {Yang}, \citenamefont {Zhang},
  \citenamefont {Song}, \citenamefont {Liu},\ and\ \citenamefont
  {Pan}}]{Han2024Jan}%
  \BibitemOpen
  \bibfield  {author} {\bibinfo {author} {\bibfnamefont {L.}~\bibnamefont
  {Han}}, \bibinfo {author} {\bibfnamefont {X.}~\bibnamefont {Fu}}, \bibinfo
  {author} {\bibfnamefont {R.}~\bibnamefont {Peng}}, \bibinfo {author}
  {\bibfnamefont {X.}~\bibnamefont {Cheng}}, \bibinfo {author} {\bibfnamefont
  {J.}~\bibnamefont {Dai}}, \bibinfo {author} {\bibfnamefont {L.}~\bibnamefont
  {Liu}}, \bibinfo {author} {\bibfnamefont {Y.}~\bibnamefont {Li}}, \bibinfo
  {author} {\bibfnamefont {Y.}~\bibnamefont {Zhang}}, \bibinfo {author}
  {\bibfnamefont {W.}~\bibnamefont {Zhu}}, \bibinfo {author} {\bibfnamefont
  {H.}~\bibnamefont {Bai}}, \bibinfo {author} {\bibfnamefont {Y.}~\bibnamefont
  {Zhou}}, \bibinfo {author} {\bibfnamefont {S.}~\bibnamefont {Liang}},
  \bibinfo {author} {\bibfnamefont {C.}~\bibnamefont {Chen}}, \bibinfo {author}
  {\bibfnamefont {Q.}~\bibnamefont {Wang}}, \bibinfo {author} {\bibfnamefont
  {X.}~\bibnamefont {Chen}}, \bibinfo {author} {\bibfnamefont {L.}~\bibnamefont
  {Yang}}, \bibinfo {author} {\bibfnamefont {Y.}~\bibnamefont {Zhang}},
  \bibinfo {author} {\bibfnamefont {C.}~\bibnamefont {Song}}, \bibinfo {author}
  {\bibfnamefont {J.}~\bibnamefont {Liu}},\ and\ \bibinfo {author}
  {\bibfnamefont {F.}~\bibnamefont {Pan}},\ }\bibfield  {title} {\bibinfo
  {title} {{Electrical 180{\ifmmode\mbox{\textdegree}\else\textdegree\fi}
  switching of N{\ifmmode\acute{e}\else\'{e}\fi}el vector in spin-splitting
  antiferromagnet}},\ }\href {https://doi.org/10.1126/sciadv.adn0479}
  {\bibfield  {journal} {\bibinfo  {journal} {Sci. Adv.}\ }\textbf {\bibinfo
  {volume} {10}},\ \bibinfo {pages} {adn0479} (\bibinfo {year}
  {2024})}\BibitemShut {NoStop}%
\bibitem [{\citenamefont {Biniskos}\ \emph {et~al.}(2025)\citenamefont
  {Biniskos}, \citenamefont {dos Santos~Dias}, \citenamefont {Agrestini},
  \citenamefont {Svit{\ifmmode\acute{a}\else\'{a}\fi}k}, \citenamefont {Zhou},
  \citenamefont
  {Posp{\ifmmode\acute{\imath}\else\'{\i}\fi}{\ifmmode\check{s}\else\v{s}\fi}il},\
  and\ \citenamefont
  {{\ifmmode\check{C}\else\v{C}\fi}erm{\ifmmode\acute{a}\else\'{a}\fi}k}}]{Biniskos2025Oct}%
  \BibitemOpen
  \bibfield  {author} {\bibinfo {author} {\bibfnamefont {N.}~\bibnamefont
  {Biniskos}}, \bibinfo {author} {\bibfnamefont {M.}~\bibnamefont {dos
  Santos~Dias}}, \bibinfo {author} {\bibfnamefont {S.}~\bibnamefont
  {Agrestini}}, \bibinfo {author} {\bibfnamefont {D.}~\bibnamefont
  {Svit{\ifmmode\acute{a}\else\'{a}\fi}k}}, \bibinfo {author} {\bibfnamefont
  {K.-J.}\ \bibnamefont {Zhou}}, \bibinfo {author} {\bibfnamefont
  {J.}~\bibnamefont
  {Posp{\ifmmode\acute{\imath}\else\'{\i}\fi}{\ifmmode\check{s}\else\v{s}\fi}il}},\
  and\ \bibinfo {author} {\bibfnamefont {P.}~\bibnamefont
  {{\ifmmode\check{C}\else\v{C}\fi}erm{\ifmmode\acute{a}\else\'{a}\fi}k}},\
  }\bibfield  {title} {\bibinfo {title} {{Systematic mapping of altermagnetic
  magnons by resonant inelastic X-ray circular dichroism}},\ }\href
  {https://doi.org/10.1038/s41467-025-64322-0} {\bibfield  {journal} {\bibinfo
  {journal} {Nat. Commun.}\ }\textbf {\bibinfo {volume} {16}},\ \bibinfo
  {pages} {9311} (\bibinfo {year} {2025})}\BibitemShut {NoStop}%
\bibitem [{\citenamefont {Das}\ \emph {et~al.}(2024)\citenamefont {Das},
  \citenamefont {Leeb}, \citenamefont {Knolle},\ and\ \citenamefont
  {Knap}}]{Das2024Jun}%
  \BibitemOpen
  \bibfield  {author} {\bibinfo {author} {\bibfnamefont {P.}~\bibnamefont
  {Das}}, \bibinfo {author} {\bibfnamefont {V.}~\bibnamefont {Leeb}}, \bibinfo
  {author} {\bibfnamefont {J.}~\bibnamefont {Knolle}},\ and\ \bibinfo {author}
  {\bibfnamefont {M.}~\bibnamefont {Knap}},\ }\bibfield  {title} {\bibinfo
  {title} {{Realizing Altermagnetism in Fermi-Hubbard Models with Ultracold
  Atoms}},\ }\href {https://doi.org/10.1103/PhysRevLett.132.263402} {\bibfield
  {journal} {\bibinfo  {journal} {Phys. Rev. Lett.}\ }\textbf {\bibinfo
  {volume} {132}},\ \bibinfo {pages} {263402} (\bibinfo {year}
  {2024})}\BibitemShut {NoStop}%
\bibitem [{\citenamefont {Roig}\ \emph {et~al.}(2024)\citenamefont {Roig},
  \citenamefont {Kreisel}, \citenamefont {Yu}, \citenamefont {Andersen},\ and\
  \citenamefont {Agterberg}}]{Roig2024Oct}%
  \BibitemOpen
  \bibfield  {author} {\bibinfo {author} {\bibfnamefont {M.}~\bibnamefont
  {Roig}}, \bibinfo {author} {\bibfnamefont {A.}~\bibnamefont {Kreisel}},
  \bibinfo {author} {\bibfnamefont {Y.}~\bibnamefont {Yu}}, \bibinfo {author}
  {\bibfnamefont {B.~M.}\ \bibnamefont {Andersen}},\ and\ \bibinfo {author}
  {\bibfnamefont {D.~F.}\ \bibnamefont {Agterberg}},\ }\bibfield  {title}
  {\bibinfo {title} {{Minimal models for altermagnetism}},\ }\href
  {https://doi.org/10.1103/PhysRevB.110.144412} {\bibfield  {journal} {\bibinfo
   {journal} {Phys. Rev. B}\ }\textbf {\bibinfo {volume} {110}},\ \bibinfo
  {pages} {144412} (\bibinfo {year} {2024})}\BibitemShut {NoStop}%
\bibitem [{\citenamefont {Leeb}\ \emph {et~al.}(2024)\citenamefont {Leeb},
  \citenamefont {Mook}, \citenamefont
  {{\ifmmode\check{S}\else\v{S}\fi}mejkal},\ and\ \citenamefont
  {Knolle}}]{Leeb2024Jun}%
  \BibitemOpen
  \bibfield  {author} {\bibinfo {author} {\bibfnamefont {V.}~\bibnamefont
  {Leeb}}, \bibinfo {author} {\bibfnamefont {A.}~\bibnamefont {Mook}}, \bibinfo
  {author} {\bibfnamefont {L.}~\bibnamefont
  {{\ifmmode\check{S}\else\v{S}\fi}mejkal}},\ and\ \bibinfo {author}
  {\bibfnamefont {J.}~\bibnamefont {Knolle}},\ }\bibfield  {title} {\bibinfo
  {title} {{Spontaneous Formation of Altermagnetism from Orbital Ordering}},\
  }\href {https://doi.org/10.1103/PhysRevLett.132.236701} {\bibfield  {journal}
  {\bibinfo  {journal} {Phys. Rev. Lett.}\ }\textbf {\bibinfo {volume} {132}},\
  \bibinfo {pages} {236701} (\bibinfo {year} {2024})}\BibitemShut {NoStop}%
\bibitem [{\citenamefont {Lieb}(1989)}]{Lieb1989Mar}%
  \BibitemOpen
  \bibfield  {author} {\bibinfo {author} {\bibfnamefont {E.~H.}\ \bibnamefont
  {Lieb}},\ }\bibfield  {title} {\bibinfo {title} {{Two theorems on the Hubbard
  model}},\ }\href {https://doi.org/10.1103/PhysRevLett.62.1201} {\bibfield
  {journal} {\bibinfo  {journal} {Phys. Rev. Lett.}\ }\textbf {\bibinfo
  {volume} {62}},\ \bibinfo {pages} {1201} (\bibinfo {year}
  {1989})}\BibitemShut {NoStop}%
\bibitem [{\citenamefont {Brekke}\ \emph {et~al.}(2023)\citenamefont {Brekke},
  \citenamefont {Brataas},\ and\ \citenamefont {Sudb{\o}}}]{Brekke2023Dec}%
  \BibitemOpen
  \bibfield  {author} {\bibinfo {author} {\bibfnamefont {B.}~\bibnamefont
  {Brekke}}, \bibinfo {author} {\bibfnamefont {A.}~\bibnamefont {Brataas}},\
  and\ \bibinfo {author} {\bibfnamefont {A.}~\bibnamefont {Sudb{\o}}},\
  }\bibfield  {title} {\bibinfo {title} {{Two-dimensional altermagnets:
  Superconductivity in a minimal microscopic model}},\ }\href
  {https://doi.org/10.1103/PhysRevB.108.224421} {\bibfield  {journal} {\bibinfo
   {journal} {Phys. Rev. B}\ }\textbf {\bibinfo {volume} {108}},\ \bibinfo
  {pages} {224421} (\bibinfo {year} {2023})}\BibitemShut {NoStop}%
\bibitem [{\citenamefont {M{\ae}land}\ \emph {et~al.}(2024)\citenamefont
  {M{\ae}land}, \citenamefont {Brekke},\ and\ \citenamefont
  {Sudb{\o}}}]{MaelandLieb24}%
  \BibitemOpen
  \bibfield  {author} {\bibinfo {author} {\bibfnamefont {K.}~\bibnamefont
  {M{\ae}land}}, \bibinfo {author} {\bibfnamefont {B.}~\bibnamefont {Brekke}},\
  and\ \bibinfo {author} {\bibfnamefont {A.}~\bibnamefont {Sudb{\o}}},\
  }\bibfield  {title} {\bibinfo {title} {{Many-body effects on
  superconductivity mediated by double-magnon processes in altermagnets}},\
  }\href {https://doi.org/10.1103/PhysRevB.109.134515} {\bibfield  {journal}
  {\bibinfo  {journal} {Phys. Rev. B}\ }\textbf {\bibinfo {volume} {109}},\
  \bibinfo {pages} {134515} (\bibinfo {year} {2024})}\BibitemShut {NoStop}%
\bibitem [{\citenamefont {Leraand}\ \emph {et~al.}(2025)\citenamefont
  {Leraand}, \citenamefont {M{\ae}land},\ and\ \citenamefont
  {Sudb{\o}}}]{Leraand2025Sep}%
  \BibitemOpen
  \bibfield  {author} {\bibinfo {author} {\bibfnamefont {K.}~\bibnamefont
  {Leraand}}, \bibinfo {author} {\bibfnamefont {K.}~\bibnamefont
  {M{\ae}land}},\ and\ \bibinfo {author} {\bibfnamefont {A.}~\bibnamefont
  {Sudb{\o}}},\ }\bibfield  {title} {\bibinfo {title} {{Phonon-mediated
  spin-polarized superconductivity in altermagnets}},\ }\href
  {https://doi.org/10.1103/g4dl-1ff2} {\bibfield  {journal} {\bibinfo
  {journal} {Phys. Rev. B}\ }\textbf {\bibinfo {volume} {112}},\ \bibinfo
  {pages} {104510} (\bibinfo {year} {2025})}\BibitemShut {NoStop}%
\bibitem [{\citenamefont {Kaushal}\ and\ \citenamefont
  {Franz}(2025)}]{franzlieblattice}%
  \BibitemOpen
  \bibfield  {author} {\bibinfo {author} {\bibfnamefont {N.}~\bibnamefont
  {Kaushal}}\ and\ \bibinfo {author} {\bibfnamefont {M.}~\bibnamefont
  {Franz}},\ }\bibfield  {title} {\bibinfo {title} {{Altermagnetism in Modified
  Lieb Lattice Hubbard Model}},\ }\href
  {https://doi.org/10.1103/PhysRevLett.135.156502} {\bibfield  {journal}
  {\bibinfo  {journal} {Phys. Rev. Lett.}\ }\textbf {\bibinfo {volume} {135}},\
  \bibinfo {pages} {156502} (\bibinfo {year} {2025})}\BibitemShut {NoStop}%
\bibitem [{\citenamefont {D{\ifmmode\ddot{u}\else\"{u}\fi}rrnagel}\ \emph
  {et~al.}(2025)\citenamefont {D{\ifmmode\ddot{u}\else\"{u}\fi}rrnagel},
  \citenamefont {Hohmann}, \citenamefont {Maity}, \citenamefont {Seufert},
  \citenamefont {Klett}, \citenamefont {Klebl},\ and\ \citenamefont
  {Thomale}}]{Durrnagel2025JulLieb}%
  \BibitemOpen
  \bibfield  {author} {\bibinfo {author} {\bibfnamefont {M.}~\bibnamefont
  {D{\ifmmode\ddot{u}\else\"{u}\fi}rrnagel}}, \bibinfo {author} {\bibfnamefont
  {H.}~\bibnamefont {Hohmann}}, \bibinfo {author} {\bibfnamefont
  {A.}~\bibnamefont {Maity}}, \bibinfo {author} {\bibfnamefont
  {J.}~\bibnamefont {Seufert}}, \bibinfo {author} {\bibfnamefont
  {M.}~\bibnamefont {Klett}}, \bibinfo {author} {\bibfnamefont
  {L.}~\bibnamefont {Klebl}},\ and\ \bibinfo {author} {\bibfnamefont
  {R.}~\bibnamefont {Thomale}},\ }\bibfield  {title} {\bibinfo {title}
  {{Altermagnetic Phase Transition in a Lieb Metal}},\ }\href
  {https://doi.org/10.1103/PhysRevLett.135.036502} {\bibfield  {journal}
  {\bibinfo  {journal} {Phys. Rev. Lett.}\ }\textbf {\bibinfo {volume} {135}},\
  \bibinfo {pages} {036502} (\bibinfo {year} {2025})}\BibitemShut {NoStop}%
\bibitem [{\citenamefont {Xu}\ and\ \citenamefont {Yang}(2025)}]{Xu2025Lieb}%
  \BibitemOpen
  \bibfield  {author} {\bibinfo {author} {\bibfnamefont {X.}~\bibnamefont
  {Xu}}\ and\ \bibinfo {author} {\bibfnamefont {L.}~\bibnamefont {Yang}},\
  }\bibfield  {title} {\bibinfo {title} {{Alterpiezoresponse in Two-Dimensional
  Lieb-Lattice Altermagnets}},\ }\href
  {https://doi.org/10.1021/acs.nanolett.5c02295} {\bibfield  {journal}
  {\bibinfo  {journal} {Nano Lett.}\ }\textbf {\bibinfo {volume} {25}},\
  \bibinfo {pages} {11870} (\bibinfo {year} {2025})}\BibitemShut {NoStop}%
\bibitem [{\citenamefont {Chang}\ \emph {et~al.}(2026)\citenamefont {Chang},
  \citenamefont {Belashchenko},\ and\ \citenamefont {Mazin}}]{Chang2025Lieb}%
  \BibitemOpen
  \bibfield  {author} {\bibinfo {author} {\bibfnamefont {P.-H.}\ \bibnamefont
  {Chang}}, \bibinfo {author} {\bibfnamefont {K.~D.}\ \bibnamefont
  {Belashchenko}},\ and\ \bibinfo {author} {\bibfnamefont {I.~I.}\ \bibnamefont
  {Mazin}},\ }\bibfield  {title} {\bibinfo {title} {{Inverse Lieb materials:
  altermagnetism and more}},\ }\bibfield  {journal} {\bibinfo  {journal} {npj
  Quantum Materials}\ }\href {https://doi.org/10.1038/s41535-026-00880-w}
  {10.1038/s41535-026-00880-w} (\bibinfo {year} {2026})\BibitemShut {NoStop}%
\bibitem [{\citenamefont {Petermann}\ \emph {et~al.}(2025)\citenamefont
  {Petermann}, \citenamefont {M{\ae}land},\ and\ \citenamefont
  {Trauzettel}}]{Petermann2025Dec}%
  \BibitemOpen
  \bibfield  {author} {\bibinfo {author} {\bibfnamefont {E.}~\bibnamefont
  {Petermann}}, \bibinfo {author} {\bibfnamefont {K.}~\bibnamefont
  {M{\ae}land}},\ and\ \bibinfo {author} {\bibfnamefont {B.}~\bibnamefont
  {Trauzettel}},\ }\bibfield  {title} {\bibinfo {title} {{Spin-resolved
  quasiparticle interference patterns on altermagnets via non-spin-resolved
  scanning tunneling microscopy}},\ }\href {https://doi.org/10.1103/sg3g-crcz}
  {\bibfield  {journal} {\bibinfo  {journal} {Phys. Rev. B}\ }\textbf {\bibinfo
  {volume} {112}},\ \bibinfo {pages} {214450} (\bibinfo {year}
  {2025})}\BibitemShut {NoStop}%
\bibitem [{\citenamefont {Maciejko}\ and\ \citenamefont
  {Rayan}(2021)}]{Maciejko2021Sep}%
  \BibitemOpen
  \bibfield  {author} {\bibinfo {author} {\bibfnamefont {J.}~\bibnamefont
  {Maciejko}}\ and\ \bibinfo {author} {\bibfnamefont {S.}~\bibnamefont
  {Rayan}},\ }\bibfield  {title} {\bibinfo {title} {{Hyperbolic band theory}},\
  }\href {https://doi.org/10.1126/sciadv.abe9170} {\bibfield  {journal}
  {\bibinfo  {journal} {Sci. Adv.}\ }\textbf {\bibinfo {volume} {7}},\ \bibinfo
  {pages} {eabe9170} (\bibinfo {year} {2021})}\BibitemShut {NoStop}%
\bibitem [{\citenamefont {Maciejko}\ and\ \citenamefont
  {Rayan}(2022)}]{Maciejko2022Mar}%
  \BibitemOpen
  \bibfield  {author} {\bibinfo {author} {\bibfnamefont {J.}~\bibnamefont
  {Maciejko}}\ and\ \bibinfo {author} {\bibfnamefont {S.}~\bibnamefont
  {Rayan}},\ }\bibfield  {title} {\bibinfo {title} {{Automorphic Bloch theorems
  for hyperbolic lattices}},\ }\href {https://doi.org/10.1073/pnas.2116869119}
  {\bibfield  {journal} {\bibinfo  {journal} {Proc. Natl. Acad. Sci. U.S.A.}\
  }\textbf {\bibinfo {volume} {119}},\ \bibinfo {pages} {e2116869119} (\bibinfo
  {year} {2022})}\BibitemShut {NoStop}%
\bibitem [{\citenamefont {Boettcher}\ \emph {et~al.}(2022)\citenamefont
  {Boettcher}, \citenamefont {Gorshkov}, \citenamefont
  {Koll{\ifmmode\acute{a}\else\'{a}\fi}r}, \citenamefont {Maciejko},
  \citenamefont {Rayan},\ and\ \citenamefont {Thomale}}]{Boettcher2022Mar}%
  \BibitemOpen
  \bibfield  {author} {\bibinfo {author} {\bibfnamefont {I.}~\bibnamefont
  {Boettcher}}, \bibinfo {author} {\bibfnamefont {A.~V.}\ \bibnamefont
  {Gorshkov}}, \bibinfo {author} {\bibfnamefont {A.~J.}\ \bibnamefont
  {Koll{\ifmmode\acute{a}\else\'{a}\fi}r}}, \bibinfo {author} {\bibfnamefont
  {J.}~\bibnamefont {Maciejko}}, \bibinfo {author} {\bibfnamefont
  {S.}~\bibnamefont {Rayan}},\ and\ \bibinfo {author} {\bibfnamefont
  {R.}~\bibnamefont {Thomale}},\ }\bibfield  {title} {\bibinfo {title}
  {{Crystallography of hyperbolic lattices}},\ }\href
  {https://doi.org/10.1103/PhysRevB.105.125118} {\bibfield  {journal} {\bibinfo
   {journal} {Phys. Rev. B}\ }\textbf {\bibinfo {volume} {105}},\ \bibinfo
  {pages} {125118} (\bibinfo {year} {2022})}\BibitemShut {NoStop}%
\bibitem [{\citenamefont {Chen}\ \emph {et~al.}(2023)\citenamefont {Chen},
  \citenamefont {Guan}, \citenamefont {Lenggenhager}, \citenamefont {Maciejko},
  \citenamefont {Boettcher},\ and\ \citenamefont
  {Bzdu{\ifmmode\check{s}\else\v{s}\fi}ek}}]{Chen2023Aug}%
  \BibitemOpen
  \bibfield  {author} {\bibinfo {author} {\bibfnamefont {A.}~\bibnamefont
  {Chen}}, \bibinfo {author} {\bibfnamefont {Y.}~\bibnamefont {Guan}}, \bibinfo
  {author} {\bibfnamefont {P.~M.}\ \bibnamefont {Lenggenhager}}, \bibinfo
  {author} {\bibfnamefont {J.}~\bibnamefont {Maciejko}}, \bibinfo {author}
  {\bibfnamefont {I.}~\bibnamefont {Boettcher}},\ and\ \bibinfo {author}
  {\bibfnamefont {T.}~\bibnamefont {Bzdu{\ifmmode\check{s}\else\v{s}\fi}ek}},\
  }\bibfield  {title} {\bibinfo {title} {{Symmetry and topology of hyperbolic
  Haldane models}},\ }\href {https://doi.org/10.1103/PhysRevB.108.085114}
  {\bibfield  {journal} {\bibinfo  {journal} {Phys. Rev. B}\ }\textbf {\bibinfo
  {volume} {108}},\ \bibinfo {pages} {085114} (\bibinfo {year}
  {2023})}\BibitemShut {NoStop}%
\bibitem [{\citenamefont {Koll{\ifmmode\acute{a}\else\'{a}\fi}r}\ \emph
  {et~al.}(2019)\citenamefont {Koll{\ifmmode\acute{a}\else\'{a}\fi}r},
  \citenamefont {Fitzpatrick},\ and\ \citenamefont {Houck}}]{Kollar2019Jul}%
  \BibitemOpen
  \bibfield  {author} {\bibinfo {author} {\bibfnamefont {A.~J.}\ \bibnamefont
  {Koll{\ifmmode\acute{a}\else\'{a}\fi}r}}, \bibinfo {author} {\bibfnamefont
  {M.}~\bibnamefont {Fitzpatrick}},\ and\ \bibinfo {author} {\bibfnamefont
  {A.~A.}\ \bibnamefont {Houck}},\ }\bibfield  {title} {\bibinfo {title}
  {{Hyperbolic lattices in circuit quantum electrodynamics}},\ }\href
  {https://doi.org/10.1038/s41586-019-1348-3} {\bibfield  {journal} {\bibinfo
  {journal} {Nature}\ }\textbf {\bibinfo {volume} {571}},\ \bibinfo {pages}
  {45} (\bibinfo {year} {2019})}\BibitemShut {NoStop}%
\bibitem [{\citenamefont {Zhang}\ \emph {et~al.}(2022)\citenamefont {Zhang},
  \citenamefont {Yuan}, \citenamefont {Sun}, \citenamefont {Sun},\ and\
  \citenamefont {Zhang}}]{Zhang2022May}%
  \BibitemOpen
  \bibfield  {author} {\bibinfo {author} {\bibfnamefont {W.}~\bibnamefont
  {Zhang}}, \bibinfo {author} {\bibfnamefont {H.}~\bibnamefont {Yuan}},
  \bibinfo {author} {\bibfnamefont {N.}~\bibnamefont {Sun}}, \bibinfo {author}
  {\bibfnamefont {H.}~\bibnamefont {Sun}},\ and\ \bibinfo {author}
  {\bibfnamefont {X.}~\bibnamefont {Zhang}},\ }\bibfield  {title} {\bibinfo
  {title} {{Observation of novel topological states in hyperbolic lattices}},\
  }\href {https://doi.org/10.1038/s41467-022-30631-x} {\bibfield  {journal}
  {\bibinfo  {journal} {Nat. Commun.}\ }\textbf {\bibinfo {volume} {13}},\
  \bibinfo {pages} {2937} (\bibinfo {year} {2022})}\BibitemShut {NoStop}%
\bibitem [{\citenamefont {Huang}\ \emph {et~al.}(2024)\citenamefont {Huang},
  \citenamefont {He}, \citenamefont {Zhang}, \citenamefont {Zhang},
  \citenamefont {Liu}, \citenamefont {Feng}, \citenamefont {Liu}, \citenamefont
  {Cui}, \citenamefont {Huang}, \citenamefont {Zhang},\ and\ \citenamefont
  {Zhang}}]{Huang2024Feb}%
  \BibitemOpen
  \bibfield  {author} {\bibinfo {author} {\bibfnamefont {L.}~\bibnamefont
  {Huang}}, \bibinfo {author} {\bibfnamefont {L.}~\bibnamefont {He}}, \bibinfo
  {author} {\bibfnamefont {W.}~\bibnamefont {Zhang}}, \bibinfo {author}
  {\bibfnamefont {H.}~\bibnamefont {Zhang}}, \bibinfo {author} {\bibfnamefont
  {D.}~\bibnamefont {Liu}}, \bibinfo {author} {\bibfnamefont {X.}~\bibnamefont
  {Feng}}, \bibinfo {author} {\bibfnamefont {F.}~\bibnamefont {Liu}}, \bibinfo
  {author} {\bibfnamefont {K.}~\bibnamefont {Cui}}, \bibinfo {author}
  {\bibfnamefont {Y.}~\bibnamefont {Huang}}, \bibinfo {author} {\bibfnamefont
  {W.}~\bibnamefont {Zhang}},\ and\ \bibinfo {author} {\bibfnamefont
  {X.}~\bibnamefont {Zhang}},\ }\bibfield  {title} {\bibinfo {title}
  {{Hyperbolic photonic topological insulators}},\ }\href
  {https://doi.org/10.1038/s41467-024-46035-y} {\bibfield  {journal} {\bibinfo
  {journal} {Nat. Commun.}\ }\textbf {\bibinfo {volume} {15}},\ \bibinfo
  {pages} {1647} (\bibinfo {year} {2024})}\BibitemShut {NoStop}%
\bibitem [{\citenamefont {Urwyler}\ \emph {et~al.}(2022)\citenamefont
  {Urwyler}, \citenamefont {Lenggenhager}, \citenamefont {Boettcher},
  \citenamefont {Thomale}, \citenamefont {Neupert},\ and\ \citenamefont
  {Bzdu{\ifmmode\check{s}\else\v{s}\fi}ek}}]{Urwyler2022Dec}%
  \BibitemOpen
  \bibfield  {author} {\bibinfo {author} {\bibfnamefont {D.~M.}\ \bibnamefont
  {Urwyler}}, \bibinfo {author} {\bibfnamefont {P.~M.}\ \bibnamefont
  {Lenggenhager}}, \bibinfo {author} {\bibfnamefont {I.}~\bibnamefont
  {Boettcher}}, \bibinfo {author} {\bibfnamefont {R.}~\bibnamefont {Thomale}},
  \bibinfo {author} {\bibfnamefont {T.}~\bibnamefont {Neupert}},\ and\ \bibinfo
  {author} {\bibfnamefont {T.}~\bibnamefont
  {Bzdu{\ifmmode\check{s}\else\v{s}\fi}ek}},\ }\bibfield  {title} {\bibinfo
  {title} {{Hyperbolic Topological Band Insulators}},\ }\href
  {https://doi.org/10.1103/PhysRevLett.129.246402} {\bibfield  {journal}
  {\bibinfo  {journal} {Phys. Rev. Lett.}\ }\textbf {\bibinfo {volume} {129}},\
  \bibinfo {pages} {246402} (\bibinfo {year} {2022})}\BibitemShut {NoStop}%
\bibitem [{\citenamefont {Liu}\ \emph {et~al.}(2023)\citenamefont {Liu},
  \citenamefont {Hua}, \citenamefont {Peng}, \citenamefont {Chen},\ and\
  \citenamefont {Zhou}}]{Liu2023Mar}%
  \BibitemOpen
  \bibfield  {author} {\bibinfo {author} {\bibfnamefont {Z.-R.}\ \bibnamefont
  {Liu}}, \bibinfo {author} {\bibfnamefont {C.-B.}\ \bibnamefont {Hua}},
  \bibinfo {author} {\bibfnamefont {T.}~\bibnamefont {Peng}}, \bibinfo {author}
  {\bibfnamefont {R.}~\bibnamefont {Chen}},\ and\ \bibinfo {author}
  {\bibfnamefont {B.}~\bibnamefont {Zhou}},\ }\bibfield  {title} {\bibinfo
  {title} {{Higher-order topological insulators in hyperbolic lattices}},\
  }\href {https://doi.org/10.1103/PhysRevB.107.125302} {\bibfield  {journal}
  {\bibinfo  {journal} {Phys. Rev. B}\ }\textbf {\bibinfo {volume} {107}},\
  \bibinfo {pages} {125302} (\bibinfo {year} {2023})}\BibitemShut {NoStop}%
\bibitem [{\citenamefont {Zhang}\ \emph {et~al.}(2023)\citenamefont {Zhang},
  \citenamefont {Di}, \citenamefont {Zheng}, \citenamefont {Sun},\ and\
  \citenamefont {Zhang}}]{Zhang2023Feb}%
  \BibitemOpen
  \bibfield  {author} {\bibinfo {author} {\bibfnamefont {W.}~\bibnamefont
  {Zhang}}, \bibinfo {author} {\bibfnamefont {F.}~\bibnamefont {Di}}, \bibinfo
  {author} {\bibfnamefont {X.}~\bibnamefont {Zheng}}, \bibinfo {author}
  {\bibfnamefont {H.}~\bibnamefont {Sun}},\ and\ \bibinfo {author}
  {\bibfnamefont {X.}~\bibnamefont {Zhang}},\ }\bibfield  {title} {\bibinfo
  {title} {{Hyperbolic band topology with non-trivial second Chern numbers}},\
  }\href {https://doi.org/10.1038/s41467-023-36767-8} {\bibfield  {journal}
  {\bibinfo  {journal} {Nat. Commun.}\ }\textbf {\bibinfo {volume} {14}},\
  \bibinfo {pages} {1083} (\bibinfo {year} {2023})}\BibitemShut {NoStop}%
\bibitem [{\citenamefont {Tummuru}\ \emph {et~al.}(2024)\citenamefont
  {Tummuru}, \citenamefont {Chen}, \citenamefont {Lenggenhager}, \citenamefont
  {Neupert}, \citenamefont {Maciejko},\ and\ \citenamefont
  {Bzdu{\ifmmode\check{s}\else\v{s}\fi}ek}}]{Tummuru2024May}%
  \BibitemOpen
  \bibfield  {author} {\bibinfo {author} {\bibfnamefont {T.}~\bibnamefont
  {Tummuru}}, \bibinfo {author} {\bibfnamefont {A.}~\bibnamefont {Chen}},
  \bibinfo {author} {\bibfnamefont {P.~M.}\ \bibnamefont {Lenggenhager}},
  \bibinfo {author} {\bibfnamefont {T.}~\bibnamefont {Neupert}}, \bibinfo
  {author} {\bibfnamefont {J.}~\bibnamefont {Maciejko}},\ and\ \bibinfo
  {author} {\bibfnamefont {T.}~\bibnamefont
  {Bzdu{\ifmmode\check{s}\else\v{s}\fi}ek}},\ }\bibfield  {title} {\bibinfo
  {title} {{Hyperbolic Non-Abelian Semimetal}},\ }\href
  {https://doi.org/10.1103/PhysRevLett.132.206601} {\bibfield  {journal}
  {\bibinfo  {journal} {Phys. Rev. Lett.}\ }\textbf {\bibinfo {volume} {132}},\
  \bibinfo {pages} {206601} (\bibinfo {year} {2024})}\BibitemShut {NoStop}%
\bibitem [{\citenamefont {Lenggenhager}\ \emph
  {et~al.}(2023{\natexlab{a}})\citenamefont {Lenggenhager}, \citenamefont
  {Maciejko},\ and\ \citenamefont
  {Bzdu{\ifmmode\check{s}\else\v{s}\fi}ek}}]{Lenggenhager2023Dec}%
  \BibitemOpen
  \bibfield  {author} {\bibinfo {author} {\bibfnamefont {P.~M.}\ \bibnamefont
  {Lenggenhager}}, \bibinfo {author} {\bibfnamefont {J.}~\bibnamefont
  {Maciejko}},\ and\ \bibinfo {author} {\bibfnamefont {T.}~\bibnamefont
  {Bzdu{\ifmmode\check{s}\else\v{s}\fi}ek}},\ }\bibfield  {title} {\bibinfo
  {title} {{Non-Abelian Hyperbolic Band Theory from Supercells}},\ }\href
  {https://doi.org/10.1103/PhysRevLett.131.226401} {\bibfield  {journal}
  {\bibinfo  {journal} {Phys. Rev. Lett.}\ }\textbf {\bibinfo {volume} {131}},\
  \bibinfo {pages} {226401} (\bibinfo {year} {2023}{\natexlab{a}})}\BibitemShut
  {NoStop}%
\bibitem [{\citenamefont {Wang}\ \emph {et~al.}(2026)\citenamefont {Wang},
  \citenamefont {Zhu},\ and\ \citenamefont {Yang}}]{Wang2026Feb}%
  \BibitemOpen
  \bibfield  {author} {\bibinfo {author} {\bibfnamefont {D.}~\bibnamefont
  {Wang}}, \bibinfo {author} {\bibfnamefont {T.}~\bibnamefont {Zhu}},\ and\
  \bibinfo {author} {\bibfnamefont {Z.}~\bibnamefont {Yang}},\ }\bibfield
  {title} {\bibinfo {title} {{Hyperbolic altermagnets with high-fold spin
  splitting}},\ }\href {https://doi.org/10.1103/PhysRevB.113.064424} {\bibfield
   {journal} {\bibinfo  {journal} {Phys. Rev. B}\ }\textbf {\bibinfo {volume}
  {113}},\ \bibinfo {pages} {064424} (\bibinfo {year} {2026})}\BibitemShut
  {NoStop}%
\bibitem [{\citenamefont {Schrauth}\ \emph {et~al.}(2023)\citenamefont
  {Schrauth}, \citenamefont {Thurn}, \citenamefont {Dusel}, \citenamefont
  {Goth}, \citenamefont {Herth},\ and\ \citenamefont {Portela}}]{hypertiling}%
  \BibitemOpen
  \bibfield  {author} {\bibinfo {author} {\bibfnamefont {M.}~\bibnamefont
  {Schrauth}}, \bibinfo {author} {\bibfnamefont {Y.}~\bibnamefont {Thurn}},
  \bibinfo {author} {\bibfnamefont {F.}~\bibnamefont {Dusel}}, \bibinfo
  {author} {\bibfnamefont {F.}~\bibnamefont {Goth}}, \bibinfo {author}
  {\bibfnamefont {D.}~\bibnamefont {Herth}},\ and\ \bibinfo {author}
  {\bibfnamefont {J.~S.~E.}\ \bibnamefont {Portela}},\ }\href
  {https://git.physik.uni-wuerzburg.de/hypertiling/hypertiling} {\bibinfo
  {title} {{hypertiling}}} (\bibinfo {year} {2023}),\ \bibinfo {note} {{version
  1.2 [git.physik.uni-wuerzburg.de/hypertiling/hypertiling]}}\BibitemShut
  {NoStop}%
\bibitem [{\citenamefont {Zhu}\ \emph {et~al.}(2025)\citenamefont {Zhu},
  \citenamefont {Huo}, \citenamefont {Feng}, \citenamefont {Zhang},
  \citenamefont {Yang},\ and\ \citenamefont {Guo}}]{Zhu2025Apr}%
  \BibitemOpen
  \bibfield  {author} {\bibinfo {author} {\bibfnamefont {X.}~\bibnamefont
  {Zhu}}, \bibinfo {author} {\bibfnamefont {X.}~\bibnamefont {Huo}}, \bibinfo
  {author} {\bibfnamefont {S.}~\bibnamefont {Feng}}, \bibinfo {author}
  {\bibfnamefont {S.-B.}\ \bibnamefont {Zhang}}, \bibinfo {author}
  {\bibfnamefont {S.~A.}\ \bibnamefont {Yang}},\ and\ \bibinfo {author}
  {\bibfnamefont {H.}~\bibnamefont {Guo}},\ }\bibfield  {title} {\bibinfo
  {title} {{Design of Altermagnetic Models from Spin Clusters}},\ }\href
  {https://doi.org/10.1103/PhysRevLett.134.166701} {\bibfield  {journal}
  {\bibinfo  {journal} {Phys. Rev. Lett.}\ }\textbf {\bibinfo {volume} {134}},\
  \bibinfo {pages} {166701} (\bibinfo {year} {2025})}\BibitemShut {NoStop}%
\bibitem [{\citenamefont {Gluscevich}\ \emph {et~al.}(2025)\citenamefont
  {Gluscevich}, \citenamefont {Samanta}, \citenamefont {Manna},\ and\
  \citenamefont {Roy}}]{Gluscevich2025Mar}%
  \BibitemOpen
  \bibfield  {author} {\bibinfo {author} {\bibfnamefont {N.}~\bibnamefont
  {Gluscevich}}, \bibinfo {author} {\bibfnamefont {A.}~\bibnamefont {Samanta}},
  \bibinfo {author} {\bibfnamefont {S.}~\bibnamefont {Manna}},\ and\ \bibinfo
  {author} {\bibfnamefont {B.}~\bibnamefont {Roy}},\ }\bibfield  {title}
  {\bibinfo {title} {{Dynamic mass generation on two-dimensional electronic
  hyperbolic lattices}},\ }\href {https://doi.org/10.1103/PhysRevB.111.L121108}
  {\bibfield  {journal} {\bibinfo  {journal} {Phys. Rev. B}\ }\textbf {\bibinfo
  {volume} {111}},\ \bibinfo {pages} {L121108} (\bibinfo {year}
  {2025})}\BibitemShut {NoStop}%
\bibitem [{\citenamefont {G{\ifmmode\ddot{o}\else\"{o}\fi}tz}\ \emph
  {et~al.}(2024)\citenamefont {G{\ifmmode\ddot{o}\else\"{o}\fi}tz},
  \citenamefont {Rein}, \citenamefont {In{\ifmmode\acute{a}\else\'{a}\fi}cio},\
  and\ \citenamefont {Assaad}}]{Gotz2024Dec}%
  \BibitemOpen
  \bibfield  {author} {\bibinfo {author} {\bibfnamefont {A.}~\bibnamefont
  {G{\ifmmode\ddot{o}\else\"{o}\fi}tz}}, \bibinfo {author} {\bibfnamefont
  {G.}~\bibnamefont {Rein}}, \bibinfo {author} {\bibfnamefont {J.~C.}\
  \bibnamefont {In{\ifmmode\acute{a}\else\'{a}\fi}cio}},\ and\ \bibinfo
  {author} {\bibfnamefont {F.~F.}\ \bibnamefont {Assaad}},\ }\bibfield  {title}
  {\bibinfo {title} {{Hubbard and Heisenberg models on hyperbolic lattices:
  Metal-insulator transitions, global antiferromagnetism, and enhanced boundary
  fluctuations}},\ }\href {https://doi.org/10.1103/PhysRevB.110.235105}
  {\bibfield  {journal} {\bibinfo  {journal} {Phys. Rev. B}\ }\textbf {\bibinfo
  {volume} {110}},\ \bibinfo {pages} {235105} (\bibinfo {year}
  {2024})}\BibitemShut {NoStop}%
\bibitem [{\citenamefont {Hosoya}(1995)}]{Hosoya1995Jun}%
  \BibitemOpen
  \bibfield  {author} {\bibinfo {author} {\bibfnamefont {H.}~\bibnamefont
  {Hosoya}},\ }\bibfield  {title} {\bibinfo {title} {{Number and shapes of the
  atomic orbitals of four and higher dimensional atoms}},\ }\href
  {https://doi.org/10.1016/0022-2860(95)08826-H} {\bibfield  {journal}
  {\bibinfo  {journal} {J. Mol. Struct.}\ }\textbf {\bibinfo {volume}
  {352-353}},\ \bibinfo {pages} {561} (\bibinfo {year} {1995})}\BibitemShut
  {NoStop}%
\bibitem [{GAP(2026)}]{GAP4}%
  \BibitemOpen
  \href {https://www.gap-system.org} {\emph {\bibinfo {title} {GAP -- Groups,
  Algorithms, and Programming, Version 4.16.0}}},\ \bibinfo {organization} {The
  GAP~Group} (\bibinfo {year} {2026})\BibitemShut {NoStop}%
\bibitem [{\citenamefont {Lenggenhager}\ \emph
  {et~al.}(2023{\natexlab{b}})\citenamefont {Lenggenhager}, \citenamefont
  {Maciejko},\ and\ \citenamefont {Bzdu{\v{s}}ek}}]{HyperCells}%
  \BibitemOpen
  \bibfield  {author} {\bibinfo {author} {\bibfnamefont {P.~M.}\ \bibnamefont
  {Lenggenhager}}, \bibinfo {author} {\bibfnamefont {J.}~\bibnamefont
  {Maciejko}},\ and\ \bibinfo {author} {\bibfnamefont {T.}~\bibnamefont
  {Bzdu{\v{s}}ek}},\ }\href {https://doi.org/10.5281/zenodo.10222598} {\bibinfo
  {title} {Hyper{C}ells: A {GAP} package for constructing primitive cells and
  supercells of hyperbolic lattices}} (\bibinfo {year}
  {2023}{\natexlab{b}})\BibitemShut {NoStop}%
\bibitem [{\citenamefont {Lenggenhager}\ \emph
  {et~al.}(2023{\natexlab{c}})\citenamefont {Lenggenhager}, \citenamefont
  {Maciejko},\ and\ \citenamefont {Bzdu{\v{s}}ek}}]{HyperBloch}%
  \BibitemOpen
  \bibfield  {author} {\bibinfo {author} {\bibfnamefont {P.~M.}\ \bibnamefont
  {Lenggenhager}}, \bibinfo {author} {\bibfnamefont {J.}~\bibnamefont
  {Maciejko}},\ and\ \bibinfo {author} {\bibfnamefont {T.}~\bibnamefont
  {Bzdu{\v{s}}ek}},\ }\href {https://doi.org/10.5281/zenodo.10222865} {\bibinfo
  {title} {{HyperBloch}: A {Mathematica} package for hyperbolic tight-binding
  models and the supercell method}} (\bibinfo {year}
  {2023}{\natexlab{c}})\BibitemShut {NoStop}%
\bibitem [{\citenamefont {Cong}\ \emph {et~al.}(2018)\citenamefont {Cong},
  \citenamefont {Savinov}, \citenamefont {Srivastava}, \citenamefont {Han},\
  and\ \citenamefont {Singh}}]{Cong2018Oct}%
  \BibitemOpen
  \bibfield  {author} {\bibinfo {author} {\bibfnamefont {L.}~\bibnamefont
  {Cong}}, \bibinfo {author} {\bibfnamefont {V.}~\bibnamefont {Savinov}},
  \bibinfo {author} {\bibfnamefont {Y.~K.}\ \bibnamefont {Srivastava}},
  \bibinfo {author} {\bibfnamefont {S.}~\bibnamefont {Han}},\ and\ \bibinfo
  {author} {\bibfnamefont {R.}~\bibnamefont {Singh}},\ }\bibfield  {title}
  {\bibinfo {title} {{A Metamaterial Analog of the Ising Model}},\ }\href
  {https://doi.org/10.1002/adma.201804210} {\bibfield  {journal} {\bibinfo
  {journal} {Adv. Mater.}\ }\textbf {\bibinfo {volume} {30}},\ \bibinfo {pages}
  {1804210} (\bibinfo {year} {2018})}\BibitemShut {NoStop}%
\bibitem [{\citenamefont {Tuz}\ \emph {et~al.}(2020)\citenamefont {Tuz},
  \citenamefont {Yu}, \citenamefont {Dmitriev},\ and\ \citenamefont
  {Kivshar}}]{Tuz2020Apr}%
  \BibitemOpen
  \bibfield  {author} {\bibinfo {author} {\bibfnamefont {V.~R.}\ \bibnamefont
  {Tuz}}, \bibinfo {author} {\bibfnamefont {P.}~\bibnamefont {Yu}}, \bibinfo
  {author} {\bibfnamefont {V.}~\bibnamefont {Dmitriev}},\ and\ \bibinfo
  {author} {\bibfnamefont {Y.~S.}\ \bibnamefont {Kivshar}},\ }\bibfield
  {title} {\bibinfo {title} {{Magnetic Dipole Ordering in Resonant Dielectric
  Metasurfaces}},\ }\href {https://doi.org/10.1103/PhysRevApplied.13.044003}
  {\bibfield  {journal} {\bibinfo  {journal} {Phys. Rev. Appl.}\ }\textbf
  {\bibinfo {volume} {13}},\ \bibinfo {pages} {044003} (\bibinfo {year}
  {2020})}\BibitemShut {NoStop}%
\end{thebibliography}
\end{document}